\documentclass[structabstract]{aa}  
\usepackage{graphicx,rotate}
\usepackage{txfonts}

\newcommand{\mtw}{$M_{20}$}

\newcommand{\re}{r_{\rm eff}}

\begin{document}

\title{AGN host galaxies at redshift z $\approx$ 0.7: peculiar or not?}

\author{A.~B\"ohm      \inst{1,2} 
\and L.~Wisotzki       \inst{2}  
\and E.~F.~Bell        \inst{3} 
\and K.~Jahnke         \inst{4} 
\and C.~Wolf           \inst{5} 
\and D.~Bacon          \inst{6} 
\and M.~Barden         \inst{1} 
\and M.~E.~Gray        \inst{7} 
\and G.~Hoeppe         \inst{8}
\and S.~Jogee          \inst{9} 
\and D.~H.~McIntosh    \inst{10} 
\and C.~Y.~Peng        \inst{11} 
\and A.~R.~Robaina     \inst{4,12}
\and M.~Balogh         \inst{13} 
\and F.~D.~Barazza     \inst{14} 
\and J.~A.~R.~Caldwell \inst{15} 
\and C.~Heymans        \inst{16} 
\and B.~H\"au\ss ler   \inst{7} 
\and E.~van~Kampen     \inst{17} 
\and K.~Lane           \inst{7} 
\and K.~Meisenheimer   \inst{4} 
\and S.~F.~S\'anchez   \inst{18} 
\and A.~N.~Taylor      \inst{16} 
\and X.~Zheng          \inst{19}
}

\offprints{Asmus.Boehm@uibk.ac.at}

\institute{Institute for Astro- and Particle Physics, University of  
Innsbruck, Technikerstr. 25/8, A-6020 Innsbruck,
   Austria
\and Astrophysikalisches Institut Potsdam, An der Sternwarte 16, D-14482 Potsdam, Germany
\and University of
Michigan, Department of Astronomy, 830 Dennison Building, 500 Church Street,
Ann Arbor, MI 48105, USA
\and Max-Planck-Institut f\"ur Astronomie,  K\"{o}nigstuhl 17, D-69117 Heidelberg, Germany   
\and Department of Physics, Denys Wilkinson Building, University of Oxford, Keble Road, Oxford, OX1 3RH, UK
\and Institute of Cosmology and Gravitation, University of Portsmouth,
Hampshire Terrace, Portsmouth, PO1 2EG, UK
\and School of Physics and Astronomy, The University of Nottingham, University Park, Nottingham NG7 2RD, UK
\and Department of Anthropology, University of Waterloo,
Waterloo, Ontario, Canada N2L 3G1
\and Department of Astronomy, University of Texas at Austin, 1 University
    Station, C1400 Austin, TX 78712-0259, USA
\and Department of Physics, University of Missouri-Kansas City, 
Kansas City, MO 64110, USA
\and Giant Magellan Telescope Organization, 251 South Lake Avenue, Pasadena, CA 91101
USA
\and Institut de Ci\'encies del Cosmos, University of Barcelona, 
08028 Barcelona, Spain
\and Department of Physics and Astronomy, University Of Waterloo, Waterloo,
Ontario, N2L 3G1, Canada
\and Department of Physics, University of Basel, 
Klingelbergstrasse 82, 4056 Basel, Switzerland
\and University of Texas, McDonald Observatory, Fort Davis, TX 79734, USA
\and Institute for Astronomy, University of Edinburgh, Blackford Hill, Edinburgh, EH9 3HJ, UK
\and European Southern Observatory, Karl-Schwarzschild-Strasse 2, D-85748 Garching bei M\"unchen, Germany 
\and Centro Hispano Aleman de Calar Alto, C/Jesus Durban Remon 2-2, E-04004 Almeria, Spain
\and Purple Mountain Observatory, National Astronomical Observatories, Chinese
Academy of Sciences, Nanjing 210008, PR China
}

\date{Received; accepted}

\abstract
{}
{We perform a quantitative morphological comparison between the hosts
of active galactic nuclei (AGN) and quiescent galaxies at
intermediate redshifts ($z \approx 0.7$). The imaging data are taken
from the large HST/ACS mosaics of the GEMS and STAGES surveys.
Our main aim is to test whether nuclear activity at this
cosmic epoch is triggered by major mergers.
}
{Using images of quiescent galaxies and stars, we created synthetic AGN
images to investigate the impact of an optical nucleus  
on the morphological analysis of AGN hosts.
Galaxy morphologies are parameterized using the 
asymmetry index $A$, the concentration 
index $C$, the Gini coefficient $G$ and the \mtw\ index. 
A sample of $\sim$\,200 synthetic AGN was matched to 21 real AGN
in terms of redshift, host brightness, and host--to--nucleus ratio
to ensure a reliable comparison between active and quiescent galaxies.
}
{
The optical nuclei strongly affect 
the morphological parameters of the underlying host galaxy.
Taking these effects into account, we find that
the morphologies of the AGN hosts are 
clearly distinct from galaxies undergoing violent
gravitational interactions. Indeed, the host galaxy
distributions in morphological
descriptor space are more similar to undisturbed galaxies than to
major mergers.
}
{Intermediate-luminosity ($L_X \la 10^{44}\,{\rm erg/s}$) AGN hosts 
at $z \approx 0.7$ show morphologies similar to the 
general population of massive galaxies with significant bulges
at the same redshifts.
If major mergers are the driver of nuclear activity at this epoch,
the signatures of gravitational interactions fade
rapidly before the optical AGN phase starts, making them undetectable 
on single-orbit HST images, at least with usual morphological 
descriptors.
This could be investigated in future synthetic observations created 
from numerical simulations of galaxy-galaxy interactions.}


\maketitle

\section{Introduction}

For a long time, active galactic nuclei (AGN) were considered a very
special class of objects, characterized by an accreting 
supermassive black hole (SMBH) in their center.
Only during the past decade, an SMBH has been found in 
practically all galaxies that are 
sufficiently close to determine the stellar kinematics in the center at high
spatial resolution and that have a significant bulge component.
It also became evident that the properties of SMBHs and their host galaxies
are coupled. 
The mass of the SMBH for instance is observed to be correlated with the 
luminosity (e.g.~Marconi \& Hunt~\cite{mar03}),
velocity dispersion (e.g.~Gebhardt et al.~\cite{geb00}; 
Ferrarese \& Merritt~\cite{fer00}), and
mass (e.g.~H\"aring \& Rix~\cite{hae04}) 
of the spheroidal component of its host galaxy.
These correlations probably point towards a fundamental link between the 
formation and evolution of galaxies and their central black holes 
(e.g., Hopkins et al.~\cite{hop08}). However, some authors have demonstrated
that the correlations might also
arise from galaxy mergers without any physically coupled growth of
black holes and galaxies (e.g.~Peng~\cite{pen07}, 
Jahnke \& Maccio~\cite{jah11}).

The main mechanism that turns a
quiescent SMBH into an AGN is still under debate.  
Candidates for this are secular processes such as bar-driven gas inflows, 
gravitational interactions, or galaxy mergers
(for reviews, see Martini~\cite{mar04}, Jogee~\cite{jog06}).
It is likely that the main triggering process will depend on the
luminosity regime and/or the cosmic epoch under investigation.

One possible approach for identifying this process is to compare the 
morphological properties of
AGN host galaxies and quiescent (non-AGN) galaxies.
To this end, some previous studies have 
used descriptors such as $CAS$ (Conselice et
al.~\cite{con00}). Grogin et al.~(\cite{gro05}) for example found no difference
between active and inactive galaxies at $0.4<z<1.3$ 
in the Chandra Deep Fields North and South.
Similarly, Pierce et al.~(\cite{pie07}) used 
the Gini coefficient $G$ and the \mtw\ index
(Lotz et al.~\cite{lot04}) to quantify the morphologies of X-ray selected AGN at
$0.2<z<1.2$ in the DEEP2 survey and concluded that the majority of the AGN
reside in nonpeculiar early-type hosts.
These studies could be taken as evidence that signs of gravitational
interactions in AGN at intermediate redshifts are scarce.
However, it should be noted that these analyses did not include a decomposition
of the AGN images into the nucleus and host components. 
Although an X-ray selection
yields a high fraction of absorbed or so-called 
optically dull AGN, it is not clear whether
optical nuclei had an impact on these results.

In a different approach, some AGN studies focused on the environment.
Serber et al.~(\cite{ser06}) have found for instance
that the galaxy density
around quasars at $z<0.4$ is enhanced with respect to
inactive $L^\ast$ galaxies.
The reported density excess is highest for the brightest quasars and 
separations $R<100$\,kpc.
Hennawi et al.~(\cite{hen06}) derived the autocorrelation function of quasars
at $0.5<z<3$ and showed an order-of-magnitude excess at $R<40\,h^{-1}$\,kpc
in comparison to the expectation from large scales $R>3\,h^{-1}$\,Mpc.
Myers et al.~(\cite{mye07}) found a lower, albeit
significant enhancement in the range $0.4<z<2.3$. 
Silverman et al.~(\cite{sil11}) compared galaxies at $0.3<z<1.1$
with and without a companion
galaxy at less than 75\,kpc separation and showed an enhanced
AGN fraction (by a factor of 2.6) in close pairs.
According to a study by Ellison et al.~(\cite{ell11}) based on 
$>$\,10$^5$ galaxies from the Sloan survey,
this result also holds in the local universe.
Although there is some disagreement on the magnitude of the density 
enhancement around AGN, these findings favor gravitational interactions or 
mergers as a triggering mechanism.

An active nucleus with significant contributions to the total flux
will bias any morphological analysis of the host galaxy in the sense that
it appears too bulge-dominated (e.g., Pierce et al.~\cite{pie10}).
Several studies have 
investigated ways to correct for the effect of a central point source 
(Sanchez et al.~\cite{san04}, 
Simmons \& Urry~\cite{sim08}, 
Kim et al.~\cite{kim08}, 
Gabor et al.~\cite{gab09},
Cisternas et al.~\cite{cis11}).
Performing a decomposition into the host and nucleus components,
Sanchez et al.~(\cite{san04}) found that 12 out of 15 
AGN at $0.5<z<1.1$ reside 
in early-type galaxies, with roughly one third of the hosts showing signs of
gravitational interactions. Moreover, the rest-frame $U-V$ colors of
the majority of the early-type hosts were significantly bluer than those of non-AGN
early-types, indicating recent or ongoing star formation.
The same has been found at similar redshifts 
for X-ray selected AGN in the Extended
Chandra Deep Field South (B\"ohm
et al.~\cite{boe07}, Silverman et al.~\cite{sil08}).
Rafferty et al.~(\cite{raf11})
covered the wide redshift range $0.3<z<3.0$ and 
showed evidence for a correlation between star formation rate and 
AGN fraction, which reaches 30\,\% for galaxies forming stars at 
$\psi \approx 1000$\,M$_\odot$/yr.
These findings could be understood in a scenario where some mechanism triggers
star formation and, potentially
with some delay, AGN activity, which eventually suppresses
star formation via feedback processes (e.g.~Schawinski et al.~\cite{sch07},
Hopkins~\cite{hop12}).

Recent studies presented evidence against major mergers being the main
triggering process: Cisternas et al.~(\cite{cis11}) for instance found no strong
distortions in $>$\,85\,\% of their sample of AGN hosts at $0.3<z<1.0$. 
Schawinski et al.~(\cite{sch11}) showed that, even
at much earlier cosmic epochs up to $z \approx 3$,
some 80\,\% of moderate-luminosity AGN (with 
$L_X < 10^{44}\,{\rm erg/s}$) reside in disk-like hosts, making a recent 
similar-mass merger very unlikely.

In our analysis, we aim to further investigate the question 
whether AGN activity is connected to strong tidal interactions. 
Our sample covers intermediate redshifts ($0.5<z<1.1$) and
X-ray luminosities 
($10^{42}\,{\rm erg/s} \lesssim L_X \lesssim 10^{44}\,{\rm erg/s}$).
We will for the first time use
real galaxy images 
for the simulation of AGN images 
\emph{and} use these for 
a quantitative
morphological comparison between quiescent and active galaxies at intermediate
redshifts. 
At variance with previous studies, our sample is not X-ray but optically
selected for AGN with broad emission lines ($\rightarrow$ type-1).

The paper is organized as follows: 
In Sect.~\ref{data}, we introduce the surveys we used and
Sect.~\ref{agnsamp} 
describes the selection of the AGN sample. 
The construction of the non-AGN comparison sample and the simulated AGN images
is outlined in Sect.~\ref{compsamp}.
We perform the morphological analysis of both samples  and discuss our results
in Sect.~\ref{morphana}; Sect.~\ref{summary} will give a short summary.
Throughout the paper, we assume a flat concordance cosmology with 
$\Omega_\Lambda=0.7$, $\Omega_m=0.3$ and $H_0=70$\,km\,s$^{-1}$\,Mpc$^{-1}$.
Unless stated otherwise, all magnitudes are given in the AB system.

\section{Data sets \label{data}}

Our morphological analysis relies on
HST/ACS imaging data taken from the GEMS survey (Rix et al. \cite{rix04}) and
the STAGES survey (Gray et al. \cite{gra09}). 

GEMS is a survey in the F606W and F850LP bands (similar to $V$ and $z$),
consisting of 78 ACS pointings. It
covers a continuous field of $28\arcmin\times 28\arcmin$ in the extended
Chandra Deep Field South. 
Details on the data reduction are given in Caldwell et al.~(\cite{cal08}). 
The STAGES survey extends over a slightly larger area 
($30\arcmin \times 30\arcmin$)
centered on the A901/2 multiple cluster system
at $z=0.165$.  With 80 tiles of 
imaging through the F606W filter, STAGES forms 
one of the largest mosaics 
ever taken with the HST.
Details on the data reduction can be found in Gray et al.~(\cite{gra09}).

In brief, the data reduction comprised the standard steps (bias subtraction,
flat-fielding, flux calibration), drizzling of the data from the original
0.05\,arcsec/pixel to a final scale of 0.03\,arcsec/pixel, background
estimation and variance determination for each pixel.
For both data sets, the
individual exposures were multi-drizzled with a Gaussian kernel
\footnote{
The initial drizzling of the GEMS data was based on a square kernel.
We here used data drizzled with a Gaussian kernel since it offers
a significantly higher sensitivity 
in detecting faint host galaxies of AGN with brighter nuclei
(Jahnke 2008, priv.~com.).}.
For the purpose of this analysis, we
used only the F606W images of GEMS, to allow a combination with 
STAGES in a homogeneous data set.

Both the STAGES and GEMS areas are  covered by COMBO-17 
(Wolf et al.~\cite{wol03}), a photometric redshift survey based on imaging in 
12 medium- and 5 broad-band filters with the Wide Field Imager at the MPG/ESO 2.2m
telescope on La Silla, Chile. COMBO-17 provides the estimated
redshifts and spectral energy
distribution (SED) classifications of several 10$^4$ objects. 
For galaxies
brighter than $R=23$ (Vega), the error on the photometric redshifts is typically
$\sigma_z \sim 0.02 (1+z)$, (Wolf et al.~\cite{wol08}). 
For non-AGN galaxies, 
stellar mass estimates $M_*$ are also available
(Wolf et al.~\cite{wol04}, Borch et al.~\cite{bor06}).  

\section{AGN sample\label{agnsamp}}

\subsection{Local point spread function \label{psf}}

The analysis of the underlying host galaxy in AGN images is a challenging
task, particularly in cases of low flux contrasts between the host and
nucleus. The key issue for accurately removing the central point source and
determining the host properties is
a good knowledge of the point spread function (PSF). 
Its width and shape 
depend on the filter, the CCD position,  the observation date, and the
spectral shape of the source. Tests on GEMS data
showed that the temporal variations are
negligible compared to the spatial variations (see Sanchez et
al.~\cite{san04}). 

Owing to the variations of the PSF shape across the CCD 
(with pixel--to--pixel flux
variations of up to 20\%), 
it is mandatory to construct individual, \emph{local} 
PSFs for each AGN.
These local PSFs were determined by normalizing and 
averaging the nearest 35 stars from all
tiles (only using stars
from GEMS for GEMS AGN and from STAGES for STAGES AGN)
around a given AGN position on the CCD. 
Very faint ($V>24$) or saturated stars were rejected.
The typical maximum distance between the AGN and the stars that were included
was $\sim$\,40\,arcsec for GEMS and $\sim$\,25\,arcsec for the
STAGES data.

A small fraction of the STAGES images (8 tiles out of 80)
were obtained $\sim$\,6 months later than the main block, which spreads
over only three weeks. 
The tiles with delayed imaging
(nos.~29, 46, 75 \dots 80; see Heymans et al.~\cite{hey08} for details)
show significant deviations of their PSFs from those of the data taken earlier
and were therefore neglected in our analysis.
To construct local PSFs just from these tiles was no option, it would
have produced PSFs with a much lower $S/N$.

\subsection{Definition of AGN sample \label{agnsample}}

Within the COMBO-17 catalog, there are 278 objects classified as 
AGN for which ACS imaging
from STAGES or GEMS is available. 
We restricted our analysis to AGN in the redshift range $0.5<z<1.1$. 
The lower limit was chosen simply due to 
the lack of AGN identified with COMBO-17 at lower redshifts. The 
upper limit is mainly motivated 
by the shift of the F606W filter's wavelength range into the rest-frame
UV, which would strongly affect the morphological appearance and limit the
comparability to objects at lower redshifts. 
Moreover, there is a dearth 
of quiescent galaxies in COMBO-17 at redshifts $z>1.1$ usable as a 
comparison sample.
Because of the SED-based selection technique, the vast majority 
of the COMBO-17 AGN are type-1, 
i.e.~they show broad emission lines and a prominent nucleus in 
optical images.  
We introduced a cut at $R<24$ Vega mag for SED and redshift reliability.
Finally, we rejected objects that were either located close to the 
edges of the tiles or observed during the delayed STAGES
observations (see previous section).
This sample comprised 28 AGN with robust photometric redshifts and 
good ACS imaging data at $0.5<z<1.1$.

As a final constraint, we checked which of the 
AGN have a detectable host galaxy.
To this end, the local PSF of each AGN was normalized to the AGN's 
flux within an
annulus of a two-pixel radius and then  subtracted from the AGN
image. By definition, this approach is an oversubtraction of the nucleus 
because the
normalization to the central flux includes the contribution of the host
galaxy. In other words, a scaled PSF subtraction provides a
conservative lower limit on the host-to-nucleus flux ratio $H/N$.
Only objects showing a flux residual of more 
than 10\% of the total flux were included in our final sample. 
This resulted in 21 AGN 
with a median redshift of $\langle z \rangle = 0.71$.

\subsection{Decomposition of AGN images \label{galfit}}

We determined the $H/N$ and the structural properties of the 
AGN host galaxies using the GALFIT package (Peng et al.~\cite{pen02}).
GALFIT allows multi-component fitting of galaxy images in a $\chi^2$
minimization approach. To decompose the AGN images, 
the individually constructed local PSF was used 
as a model for the unresolved nucleus of the
AGN, and a S\'ersic profile (S\'ersic \cite{ser68}) as a model for the host galaxy.
This model can be used to approximate various types of 
surface brightness profiles with a S\'ersic index
$n=1$ being equivalent to an exponential disk
and $n = 4$ to a de Vaucouleurs profile.

The two-component GALFIT modeling introduces a total of ten free
parameters. For the S\'ersic component, these were the position $x_{\rm s}$, 
$y_{\rm s}$, total brightness ${\rm v}_{\rm host}$,
effective radius $\re$, S\'ersic index $n$, position angle $\theta$, and axis
ratio $q$. For the nucleus model, only position $x_{\rm n}$, $y_{\rm n}$ and 
brightness 
${\rm v}_{\rm nuc}$ were free parameters. The user has to define initial
guesses for all parameters, which GALFIT uses to start the fitting process.
These initial guess values of $m_{\rm s}$, $x_{\rm s}$, $y_{\rm s}$,  $\re$, $\theta$, 
$m_{\rm n}$, $x_{\rm n}$, $y_{\rm n}$ were determined with
 Source Extractor (Bertin \&
Arnouts \cite{ba96}). 
The flux ratio between host and nucleus was set according to the
estimate from scaled PSF subtraction.
The initial guess value of the S\'ersic index was $n=2.5$, representing an
intermediate type between an $n=1$ exponential disk and a 
de Vaucouleurs profile.
Fits were performed on thumbnails of size 256\,$\times$\,256\,pixel$^2$,
corresponding to 7.68\,$\times$\,7.68\,arcsec$^2$. 
Neighboring objects and bad pixels were masked.

GALFIT allows one to set constraint intervals on the free fitting parameters. 
We required the relative positions $\Delta x = x_{\rm s} - x_{\rm n}$ and 
$\Delta y = y_{\rm s} - y_{\rm n}$ to be within five pixels, constrained the
S\'ersic index to $0.25 \le n \le 8$ and the effective radius
to 0.1\,arcsec $\le \re \le$ 3\,arcsec. All fits were visually checked, in
particular for the fit residuals. For a few objects, it was necessary to
set additional constraints, e.g.~on the position of the components, to acquire
the best fitting results.

We find median values of 
$\langle {\rm v}_{\rm host} \rangle = 23.05$,
$\langle H/N \rangle = 0.63$, 
$\langle n \rangle = 1.51$ and
$\langle \re \rangle = 0.22$\,arcsec corresponding to 
1.6\,kpc at the average redshift 
$\langle z \rangle = 0.71$ of the sample.
Host parameter distributions and a test 
of their reliability are discussed in the following section.

\subsection{X-ray luminosities \label{xray}}

Our spectrophotometric classification rests on the COMBO-17 data. 
By definition, all objects identified as AGN in this survey
(in the rest-frame UV/optical) are type-1, i.e. unabsorbed AGN.
It is not clear whether our sample covers the same range
in X-ray luminosities as \emph{X-ray-selected} samples at similar redshifts.
Nuclear activity of different luminosity regimes might be triggered
by different processes (e.g. secular evolution vs.~gravitational 
interactions), and this could complicate
a comparison between our study and others. 
This section presents an estimate of the X-ray luminosities for our sample.

Ten objects in our sample are covered by the Chandra Deep Field South
(CDFS) point source catalog 
(Lehmer et al.~\cite{leh05}). Their hardness ratios 
between the 0.5-2\,keV and the 2-8\,keV band
(${\rm HR} = ({\rm H}-{\rm S}) / ({\rm H}+{\rm S})$, 
where H and S represent the hard and soft
bands, respectively)
cover the range $-0.64<{\rm HR}<-0.40$, confirming that they are unabsorbed.
Adopting an X-ray power law slope of $\Gamma=2$, we derived the
X-ray luminosity at 2-8 keV. They span values between
$L_X=1.7 \times 10^{42}\,{\rm erg/s}$ and 
$L_X=3.2 \times 10^{44}\,{\rm erg/s}$ with
a median of $\langle L_X \rangle = 2.1 \times 10^{43}\,{\rm erg/s}$.
For the other 11 objects, we had to apply an indirect estimate. Using
the AGN template by Berk et al.~(\cite{ber01}), we
transformed the observed $v$-band nuclear brightnesses into the rest-frame
$B$-band luminosities. These, in turn, were computed into X-ray luminosities
following Hopkins et al.~(\cite{hop07}).
They cover the range 
$5.9 \times 10^{41}\,{\rm erg/s}<L_X<6.3 \times 10^{43}\,{\rm erg/s}$ with
a median of $\langle L_X \rangle = 7.2 \times 10^{42}\,{\rm erg/s}$.
We hence find that, even though the sample consists of optically selected type-1
AGN with prominent nuclear sources, most of their X-ray luminosities are 
in the intermediate range. Only 3 out of 21 are powerful AGN with 
$L_X>10^{44}\,{\rm erg/s}$.
As a test, we used the approach of Hopkins et al.~also to estimate the 
luminosities of the ten AGN covered by the CDFS. 
The indirectly computed X-ray luminosities differ from
the CDFS-based ones by only $\sim$\,0.22\,dex on average.

We now compare the luminosity range of our sample to X-ray-selected
studies at similar cosmic epochs.
Grogin et al.~(\cite{gro05}) only gave a lower limit of
$L_X > 10^{42}\,{\rm erg/s}$ 
at the high-redshift end of their $0.4<z<1.3$ sample.
The data sets of 
Pierce et al.~(\cite{pie07}, $0.2<z<1.2$), 
Gabor et al.~(\cite{gab09}, $0.3<z<1.0$), 
and Cisternas et al.~(\cite{cis11}, $0.3<z<1.0$)
span the ranges
$10^{41}\,{\rm erg/s} \la L_X \la 10^{44}\,{\rm erg/s}$, 
$10^{42}\,{\rm erg/s} \la L_X \la 4 \times 10^{44}\,{\rm erg/s}$
and
$2 \times 10^{42}\,{\rm erg/s} \la L_X \la 3 \times 10^{44}\,{\rm erg/s}$,
respectively.
Despite of the differences in selection, our sample hence covers similar
luminosities as these previous studies.

\section{Comparison sample \label{compsamp}}

In contrast to many previous studies 
we aim to take into
account the impact of the optical nucleus on the quantitative 
AGN host galaxy properties. Thus we have constructed a special kind of 
comparison sample in three main steps.

First, we defined a set of non-AGN galaxies that is matched to the
redshifts and magnitudes of our sample of 21 AGN hosts. Secondly, 
we created \emph{synthetic} AGN images
from these quiescent galaxies
that cover the same $H/N$ as the
real AGN  (section~\ref{synthagn}). 
Finally, the synthetic AGN images were decomposed into the host and nucleus
components using GALFIT in the same way as the real AGN.
After subtracting
the model nucleus from the AGN image, the
recovered host galaxies of the synthetic AGN
were combined with the host galaxies of the real AGN and 
analyzed in concentration, asymmetry, Gini and \mtw\ index
(see Section~\ref{morphana}).

\subsection{Selection}

The basis for the selection of the comparison sample was the COMBO-17
 catalog of
the GEMS field.
For the pre-selection, the following constraints were used:
\begin{enumerate}
\item Apparent Vega magnitude $R<24$. In this magnitude regime, we can rely on
  photometric redshifts accurate to within 3-4\% and robust 
SED classifications from COMBO-17 (Wolf
  et al.~\cite{wol03}).
\item Source classified as a ``galaxy'' on the basis of its SED 
(but  \emph{not} as an AGN).
\item No X-ray counterpart in the point source catalog of the Extended
  Chandra Deep Field South (Lehmer et al.~\cite{leh05}). This criterion was
  applied to exclude absorbed or ``optically dull'' AGN.
\item S\'ersic index $n>1.5$ (derived on the F850LP images, i.e.~$z$-band, 
by H\"aussler et al.~\cite{hau07}). 
This constraint was set to restrict the sample to galaxies with
significant bulge components.
According to the relation between bulge mass and SMBH mass, 
host galaxies with small bulges or even without a bulge
are not expected to be present in the AGN sample.

\end{enumerate}

These criteria were fulfilled by 3536 galaxies with GEMS data.
In the next step, ten quiescent counterparts were selected for each of the 21 AGN
hosts within a bin $\Delta z/z \le 0.025$ of the AGN redshift 
and within $\Delta {\rm v}_{\rm host} \le 0.05$\,mag 
of the total brightness of the host as
determined with the GALFIT decomposition. 
When less than 10 non-AGN were found in the catalog within these limits, 
the limits were slightly relaxed and the search
repeated iteratively until ten counterparts could be selected.
After visual inspection of all 210 objects in the resulting sample,
seven galaxies had to be rejected because they were unresolved.
The final non-AGN sample hence contained 203 galaxies.

\begin{figure}
\resizebox{\hsize}{!}{\includegraphics[angle=270]{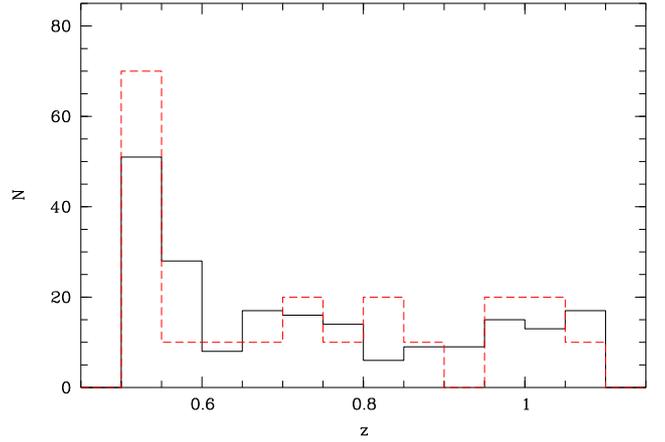}}
\caption{\label{zhis}
Comparison between the redshifts of the AGN (dashed) and the 
sample of quiescent galaxies (solid).
The histograms are normalized to the same
total number of objects.
}
\end{figure}

\begin{figure}
\resizebox{\hsize}{!}{\includegraphics[angle=270]{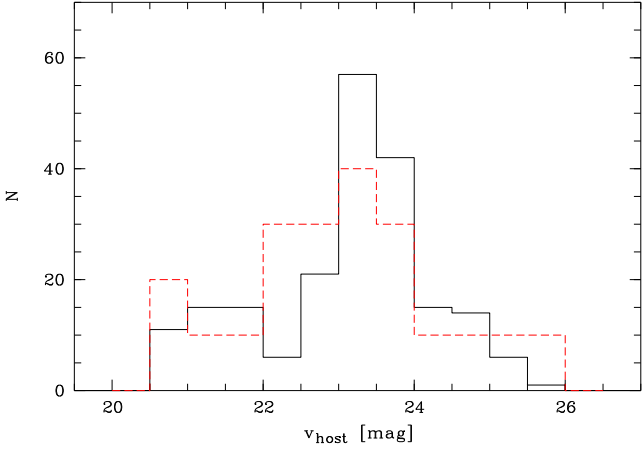}}
\caption{\label{vhis}
Distribution of the AGN host galaxy magnitudes derived with GALFIT 
(dashed) compared to the
quiescent sample (solid).
The histograms are normalized to the same
total number of objects.
}
\end{figure}

The COMBO photometric 
redshift distributions of the 21 AGN hosts and the 203 quiescent galaxies
are shown in Fig.~\ref{zhis}.  The
$z$-distribution of the AGN was normalized to a total of 203 objects
to allow a direct comparison between the two samples.
The median photometric redshift of the AGN sample is 
$\langle z \rangle = 0.71$, virtually identical to the
$\langle z \rangle = 0.70$ of the quiescent sample.
Fig.~\ref{vhis} displays the distributions in total apparent $V$ magnitude of
the AGN hosts (from GALFIT) and the quiescent comparison sample
(Source Extractor MAG\_AUTO values). Again, both samples' distributions are
normalized to the same number of objects. 
The median total magnitudes of the two samples are
$\langle {\rm v}_{\rm host}  \rangle = 23.05$ (AGN) and 
$\langle {\rm v}_{\rm host}  \rangle = 23.33$ (comparison
sample). The slight difference can be explained with the
relaxation of the constraints partly necessary during the
selection of the comparison sample.

%
%

%
%

\subsection{Visual classification\label{vis}}


%
%
\begin{figure*}
\resizebox{\hsize}{!}{\includegraphics[angle=0]{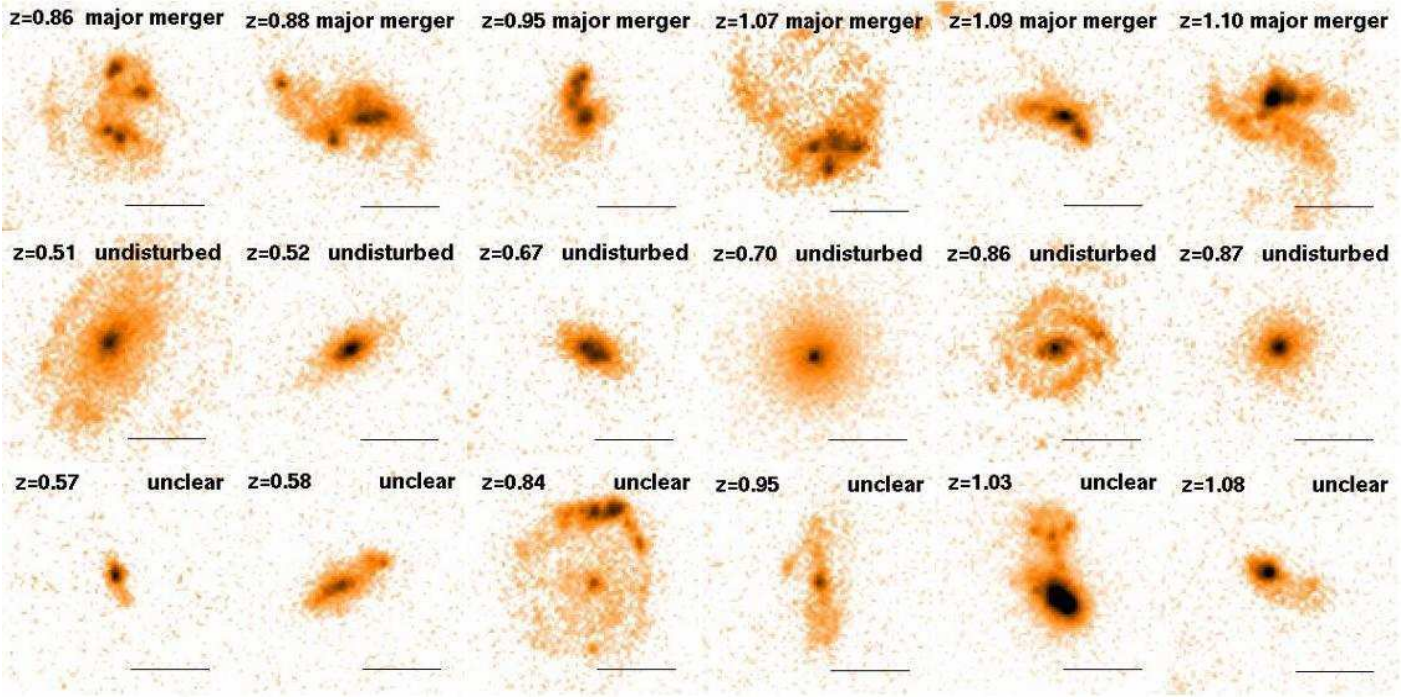}}
\caption{\label{visclass}
HST/ACS $v$-band images of
galaxies from the non-AGN comparison sample. Each row displays six
examples of the three different morphological types that were denoted via
\emph{visual classification} (with at least four out of eight
classifiers agreeing). 
The bar in each image has a length of one arcsec,
corresponding to $\sim$\,6\,kpc at $z=0.5$ or
$\sim$\,8\,kpc at $z=1$.
}
\end{figure*}

To calibrate the following quantitative 
morphological analysis, 
we performed a visual classification of all
galaxies in the quiescent sample. 
We decided to use a very plain scheme.
\emph{ Its main purpose is
to test whether major mergers are the triggering
mechanism of AGN at redshifts $0.5<z<1.1$}. 
To this end, we distinguished between three
morphological classes:

1. \emph{Major mergers}: galaxies that are heavily distorted 
(``train-wrecks"), show two nuclei of similar luminosity, or
two pronounced tidal tails of equal length.

2. \emph{Undisturbed} galaxies: no signs of recent or ongoing gravitational
interactions.

3. \emph{Unclear} cases: all other objects, including
minor mergers, ambiguous major merger candidates, or morphologies 
on the edge between undisturbed and mildly disturbed.

Using this simple scheme, all 203 galaxies in the comparison sample
were visually inspected and classified by eight of the authors 
(AB, DB, MaB, MG, GH, SJ, DM, AR). 
Comparing the individual classifications,
agreement was found
between four classifiers or more in 88\% of the cases;
between five or more in 84\% of the cases;
between six or more in 67\% of the cases,
and between seven or more in 48\% of the cases.

The classifications were merged with a constraint of at least four classifiers
agreeing on a certain type.
Cases with less than four identical classifications were
considered unclear. If an object was classified as a major merger
four times and as unclear four times,
it was also defined as unclear. We refer to the criterion of at least  
four identical classifications per object as the 4/8 sample.
Examples are presented in Fig.~\ref{visclass}.
Similarly, a scheme 5/8 requires that five
or more classifiers agree (or else, the given object is 
unclear); and equivalently for the 6/8 and 7/8 samples.
The respective number of major mergers, undisturbed, and unclear
objects are summarized in Table~\ref{tabno}.

\begin{table}[t] 
\caption{\label{tabno}Visual classifications of the 203 quiescent
comparison galaxies.}
\begin{center}
\begin{tabular}{cccc} \hline\hline
 & & & \\
Criterion & Maj. Merg. & Undist.  & Unclear \\ 
(1) & (2) & (3) & (4) \\
 & & & \\ \hline
 & & & \\
4/8 & 15  & 95 & 93 \\
5/8 & 14  & 95 & 94 \\
6/8 & 8   & 86 & 109 \\
7/8 & 4   & 68 & 131 \\ 
 & & &  \\ \hline
\end{tabular}
\end{center}
\emph{Note:} Col.~(1) denotes the required
minimum number of classifiers agreeing on the type of
a given object.
Cols.~(2) to (4) compare the respective numbers of
major mergers, undisturbed and unclear objects.
\end{table}

We tested how well the various combined classifications
 distinguish between undisturbed galaxies and
major mergers on the basis of morphological descriptors such as
concentration $C$ and asymmetry $A$,
to choose a scheme for our final analysis.
 This is described in
Sect.~\ref{finana}.

\subsection{Synthetic AGN images \label{synthagn}}

The modifications of the quiescent galaxy ACS images described in the
following constitute the backbone of our analysis. 
The same
kind of residuals that occur after subtracting the nucleus in the
central part of an AGN host were introduced in the quiescent galaxy
images. 
This ensures a 
\emph{direct comparability} 
of the morphologies of AGN hosts and non-AGN galaxies.
We achieved this by adding a \emph{synthetic nucleus} 
to a given quiescent galaxy, constructing its local PSF and 
finally decomposing it into the host and nucleus component with GALFIT. 
Thus, we analyzed synthetic and real AGN in exactly the same way.
For the synthetic nuclei, we used
real stellar images from GEMS.
It is hence ensured that the deviations between the light profile
shape of a nucleus
and the corresponding local PSF will be statistically the same as in the case
of the real AGN, and the residuals of the nucleus in simulated AGN and
real AGN images~--- after subtracting the best-fit model nucleus~---
will have the same noise properties.

The stars we used as synthetic nuclei were selected from the
GEMS catalog, constrained to
the $v$-band magnitude range of the nuclei of the real AGN, $19.6<v<25.3$.
Since the PSF shape is sensitive to the SED of a source and the nuclei of
AGN have very blue colors, we set 
color cuts of $(B-V)<0.6$ and $(V-R)<0.6$  
(both Vega magnitudes). 
Five stars were rejected because they showed
cosmics or bad pixels close to the central peak. The final list for 
creating synthetic nuclei consisted of 58 stars with median colors of
$\langle B-V \rangle \approx 0.44$ and
$\langle V-R \rangle \approx 0.43$ (Vega magnitudes).

For each of the quiescent galaxies, a star was selected 
from this list such that
the flux ratio between galaxy and star was the same as the $H/N$ of the
AGN to which the quiescent galaxy initially was assigned in the construction
of the comparison sample. Small normalizations of 
about a few percent were applied to the stellar fluxes to achieve
exactly the same flux ratios between a galaxy and the star on the one hand
and AGN host and nucleus on the other. 
After normalization, the quiescent galaxy (synthetic host) 
and stellar image (synthetic nucleus) were simply
co-added with the star positioned at the galaxy's central coordinates according to the GALFIT
modeling on the initial image.
The variance frames of the two components were
also combined to use them in the subsequent GALFIT decomposition.

\begin{figure}
\resizebox{\hsize}{!}{\includegraphics[angle=0]{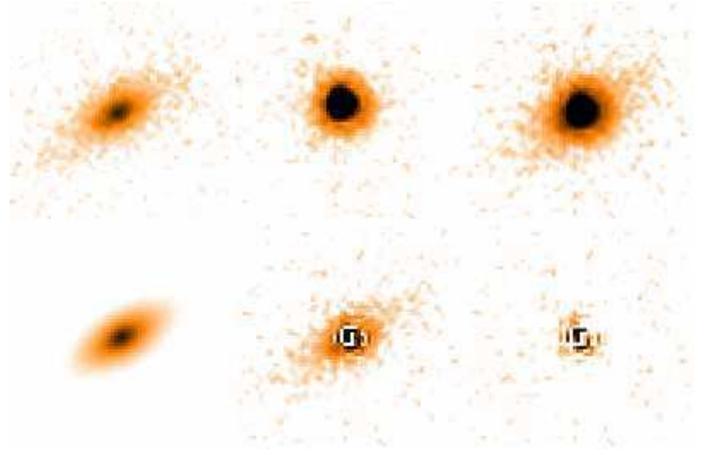}}
\caption{\label{simul}
Illustration of the AGN simulation procedure.
{\it Upper left}: original quiescent galaxy 
at redshift $z=0.52$, apparent brightness 
$v=23.27$.
{\it Upper middle}: star of brightness $v=21.46$ used as a synthetic nucleus.
{\it Upper right}: simulated AGN (co-added galaxy and star) with 
a host--to--nucleus ratio of $H/N=0.18$.
{\it Lower left}: GALFIT model of the host galaxy (from fitting a nucleus $+$
host model to the simulated AGN).
{\it Lower middle}: recovered host galaxy after subtracting the GALFIT model
of the nucleus from the simulated AGN image.
{\it Lower right}: residual after subtracting the combined galaxy $+$ nucleus
model from the simulated AGN image.
Each image is $\sim$\,2\,arcsec on the side.
}
\end{figure}

The local PSF for each synthetic AGN was created based on the CCD position of
the \emph{star} that was used as a synthetic nucleus. 
The 35 nearest stars around the CCD position of a given
synthetic nucleus were normalized and averaged. Local PSFs for the synthetic
AGN hence have the same quality as those of the real AGN.
Finally, using GALFIT, all synthetic AGN were decomposed into
nucleus and S\'ersic host with the same configuration as described for
the real AGN (section~\ref{galfit}).
The main steps of the creation and decomposition of the synthetic AGN are
illustrated in Fig.~\ref{simul}.

%
%

%
%
\begin{figure}
\resizebox{\hsize}{!}{\includegraphics[angle=270,bb=54 44 564 772]{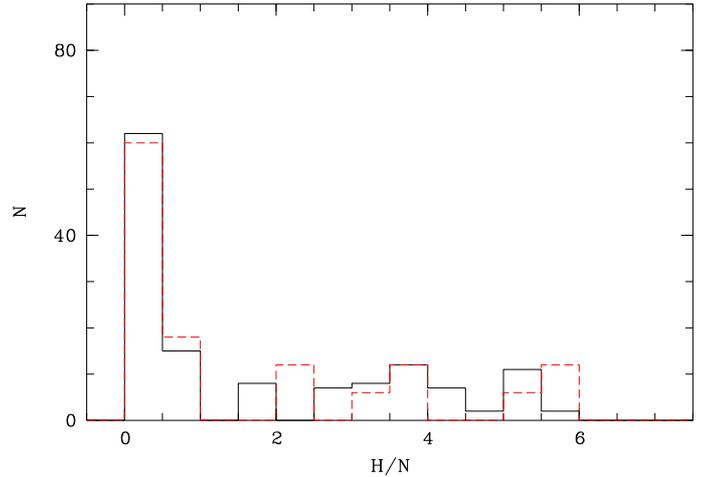}}
\caption{\label{htonhis}
Distribution of the host--to--nucleus ratios $H/N$ of the real AGN (dashed) 
compared to the simulated AGN (solid). 
The histograms are normalized to the same total number of objects.
}
\end{figure}

The $H/N$ values of the real AGN and the synthetic AGN
are compared to each other in Fig.~\ref{htonhis}. 
As in the case of the redshifts and host magnitudes, the 
two samples differ only slightly in $H/N$, 
with median values of $\langle H/N \rangle$ = 0.57 and 
$\langle H/N \rangle$ = 0.63 for the simulated AGN and  real AGN, 
respectively.
These deviations do not stem from the selection process itself but
from the limited accuracy of the GALFIT decomposition; we discuss this
in detail in the next section. 

\begin{figure*}
\resizebox{\hsize}{!}{\includegraphics[angle=0]{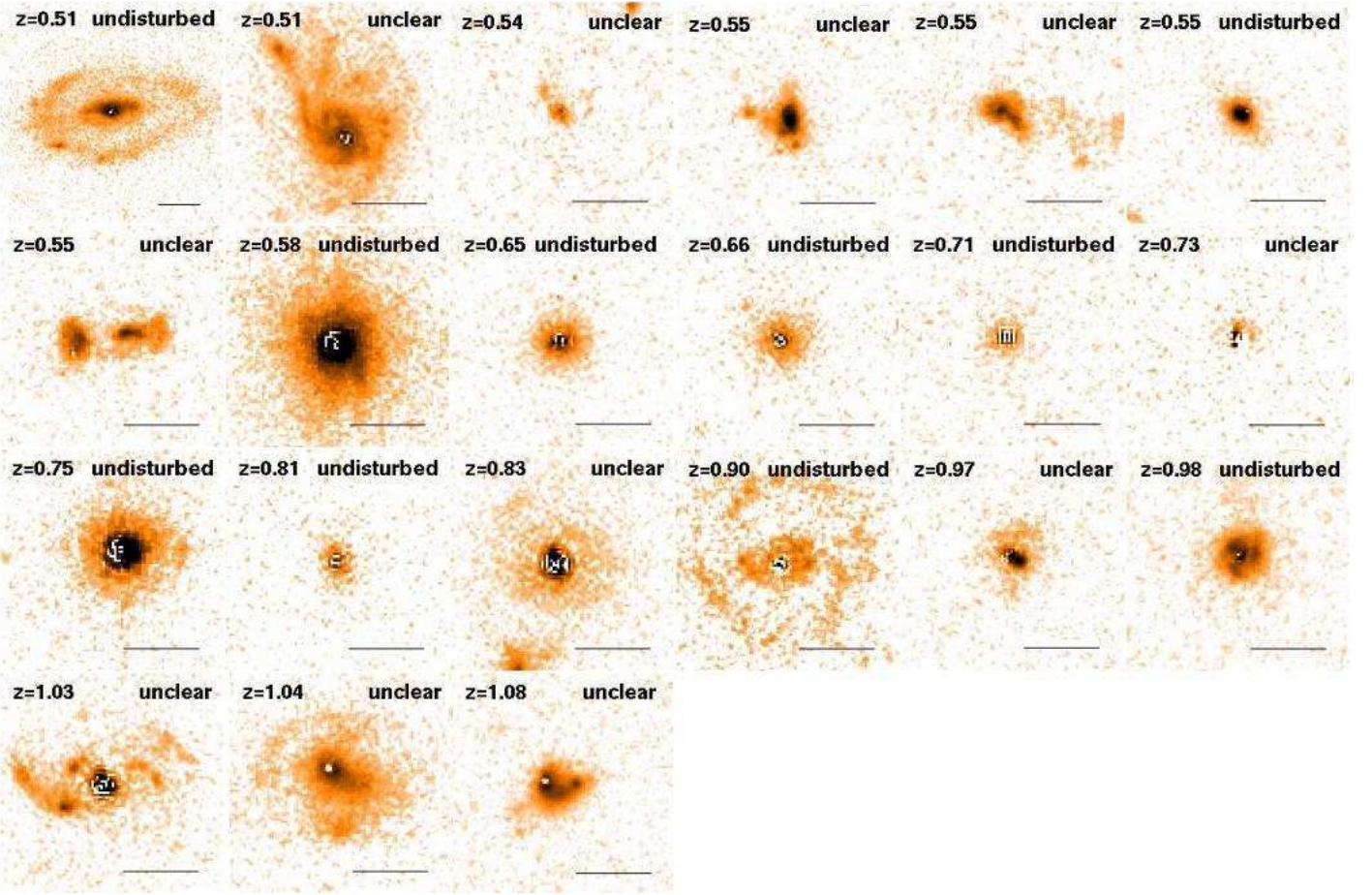}}
\caption{\label{visclass2}
Visual classifications of the nucleus-subtracted real AGN sample 
(with subtracted model nucleus from two-component modeling as 
described in Sect.\ref{galfit}). 
The bar in each image has a length of one arcsec.
}
\end{figure*}

The visual classifications described in the previous section were
also applied to the real AGN sample after subtracting the best-fit
models of the nuclear point sources from the images.
These classifications were made as a blind test with 
all nucleus-subtracted real AGN and all nucleus-subtracted simulated AGN 
randomly mixed. 
We present an overview in Table~\ref{tabagn}.
It is striking that none of the real AGN is classified as a major merger.
This is discussed in more detail in Sect.~\ref{discus}, in combination with
the results from our quantitative morphological analysis.
All real AGN and their respective visual classes 
(based on the 4/8 criterion) are shown in Fig.\ref{visclass2}.

%
%

\begin{table}[t] 
\caption{\label{tabagn}Visual classifications of the 21 real AGN hosts.}
\begin{center}
\begin{tabular}{cccc} \hline\hline
 & & & \\
Criterion & Maj. Merg. & Undist. & Unclear \\ 
(1) & (2) & (3) & (4) \\
 & & & \\ \hline
 & & & \\
4/8 & 0  & 10 & 11 \\
5/8 & 0  & 9 & 12 \\
6/8 & 0  & 7 & 14 \\
7/8 & 0  & 3 & 18 \\ 
 & & &  \\ \hline
\end{tabular}
\end{center}
\emph{Note:} Columns are the same as in Table~\ref{tabno}.
\end{table}

\subsection{Stability of the AGN decomposition}

Our synthetic AGN images can be used to test 
the GALFIT recovery of the main host galaxy parameters such as total brightness, 
effective
radius, and S\'ersic index for a broad range of $H/N$.
Similar tests have been carried out by Sanchez et al.~(\cite{san04}), 
Kim et al.~(\cite{kim08}),  and 
Gabor et al.~(\cite{gab09}); but in these studies, no \emph{real} 
galaxy images but simulated, perfectly
symmetric hosts have been used. Only Simmons \& Urry~(\cite{sim08}) used
observed data for part of their tests. 

\begin{figure*}
\resizebox{\hsize}{!}{\includegraphics[height=8.5cm,angle=270]{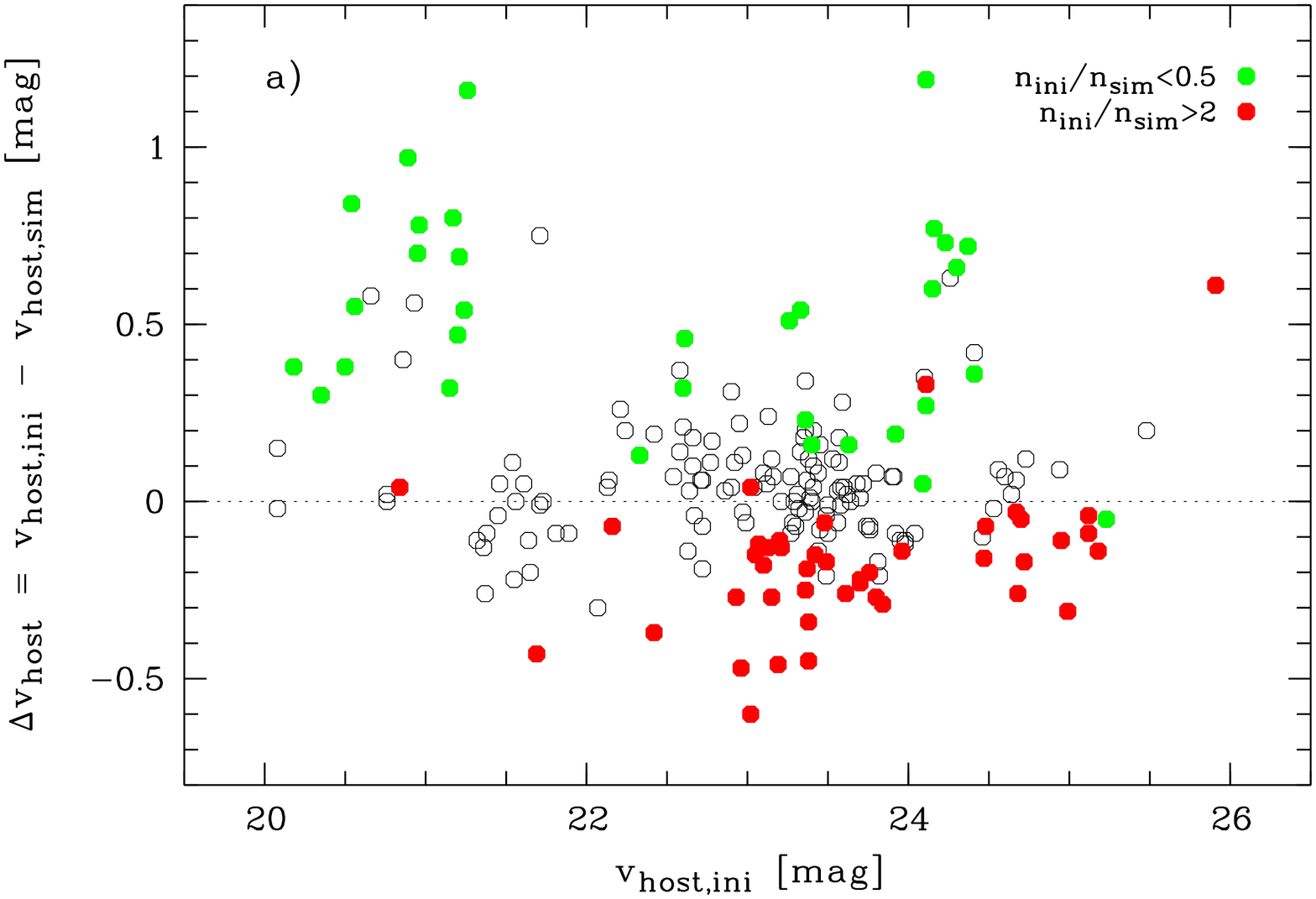}
\includegraphics[height=8.5cm,angle=270]{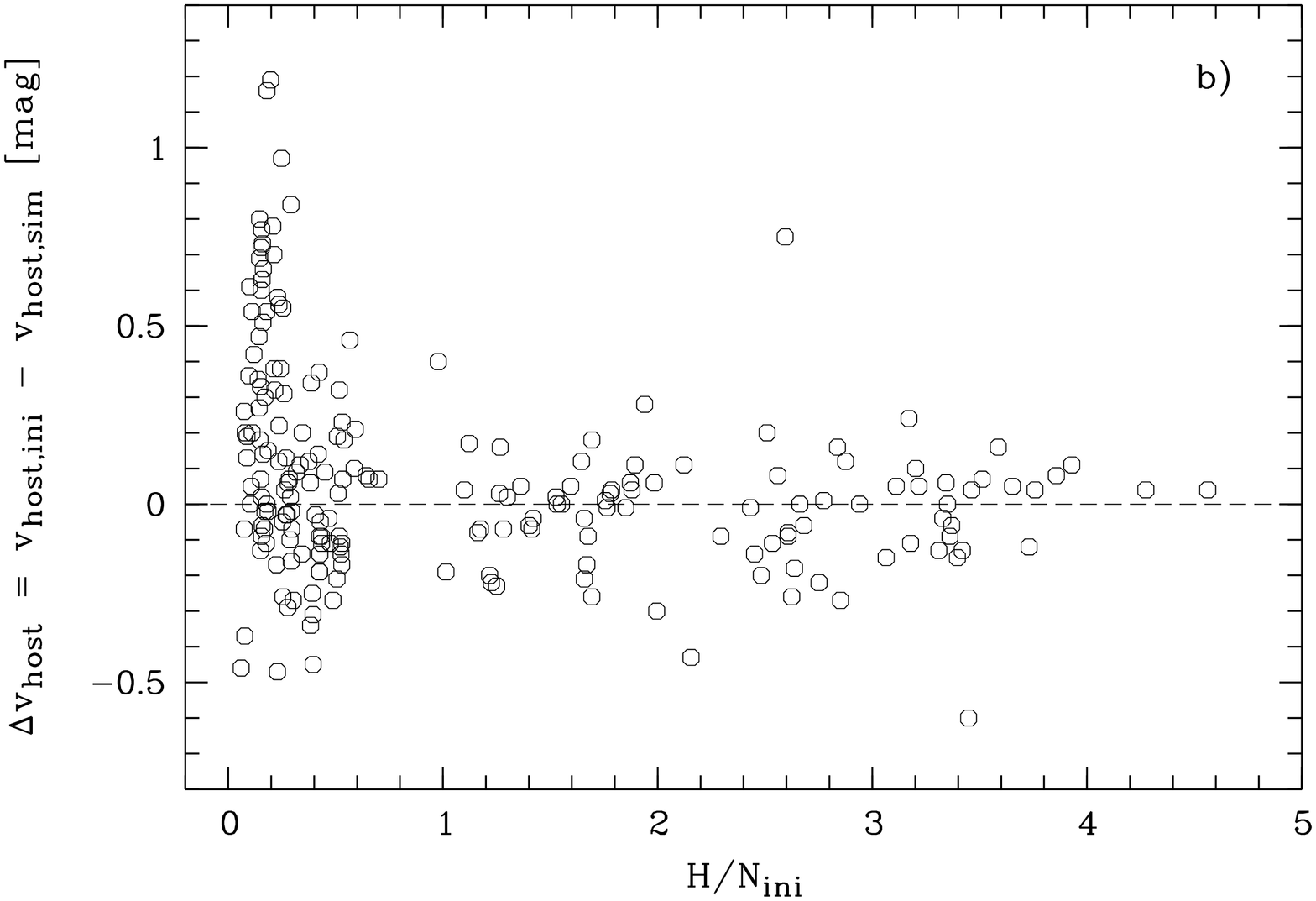}}
\resizebox{\hsize}{!}{\includegraphics[height=8.5cm,angle=270]{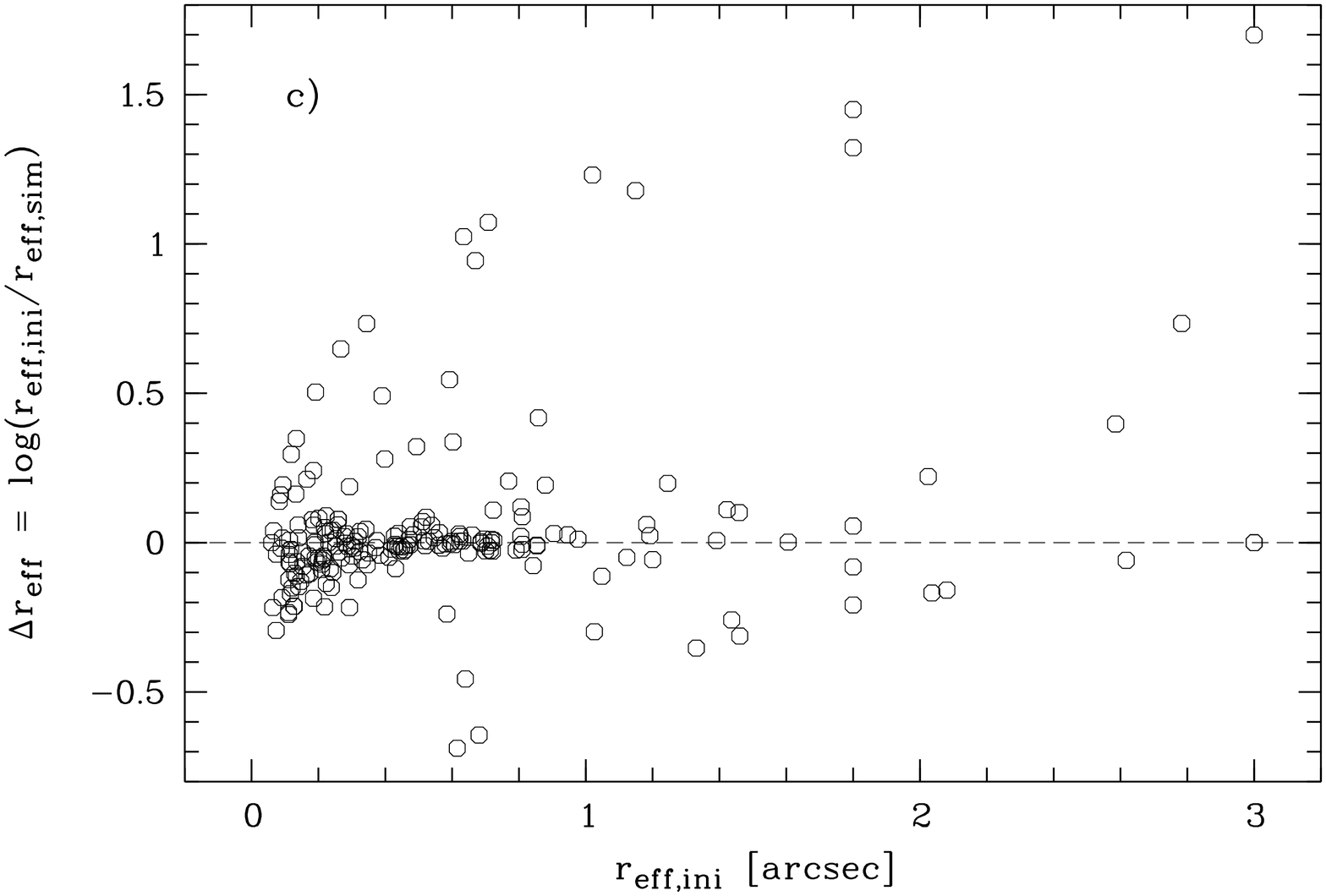}
\includegraphics[height=8.5cm,angle=270]{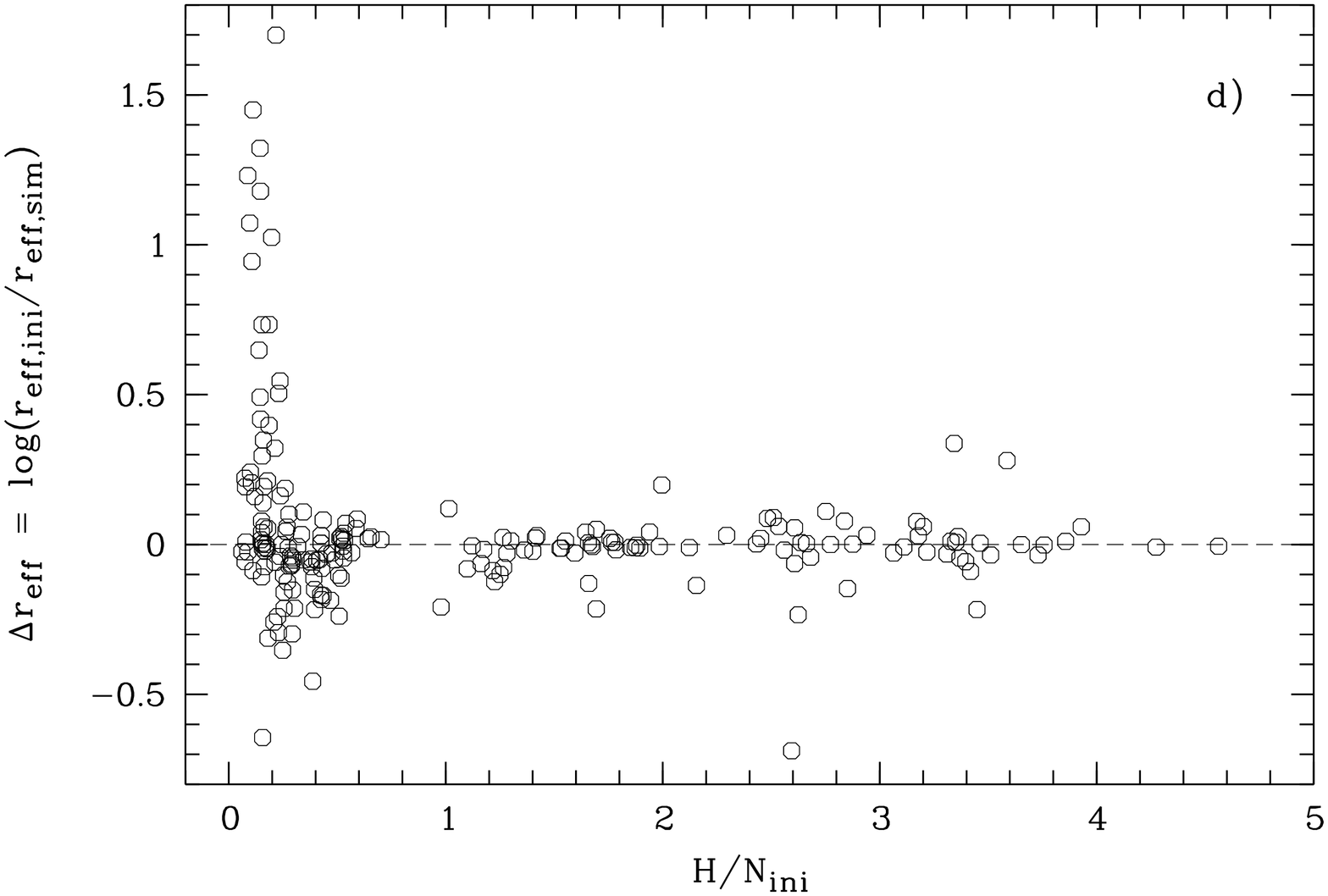}}
\resizebox{\hsize}{!}{\includegraphics[height=8.5cm,angle=270]{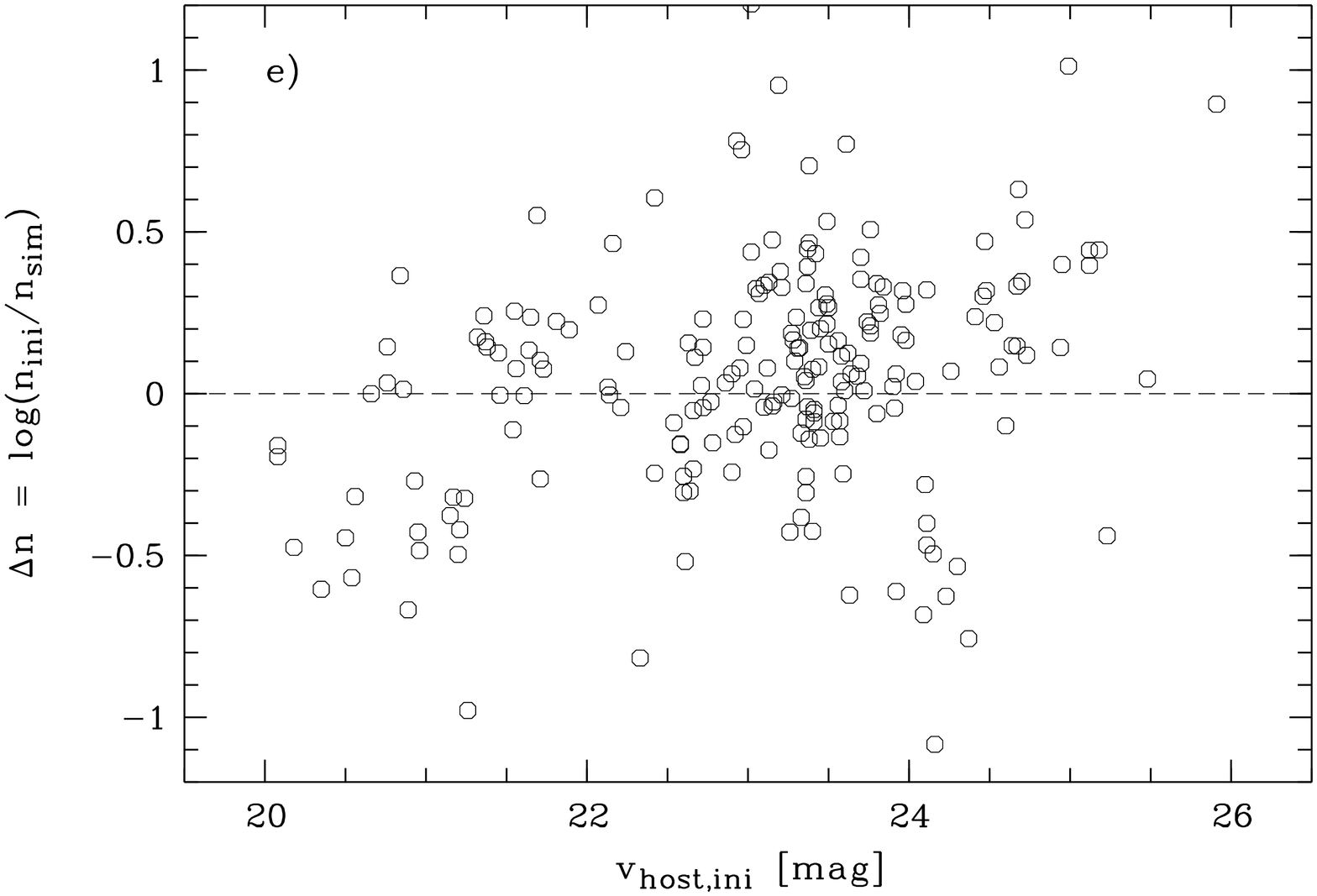}
\includegraphics[height=8.5cm,angle=270]{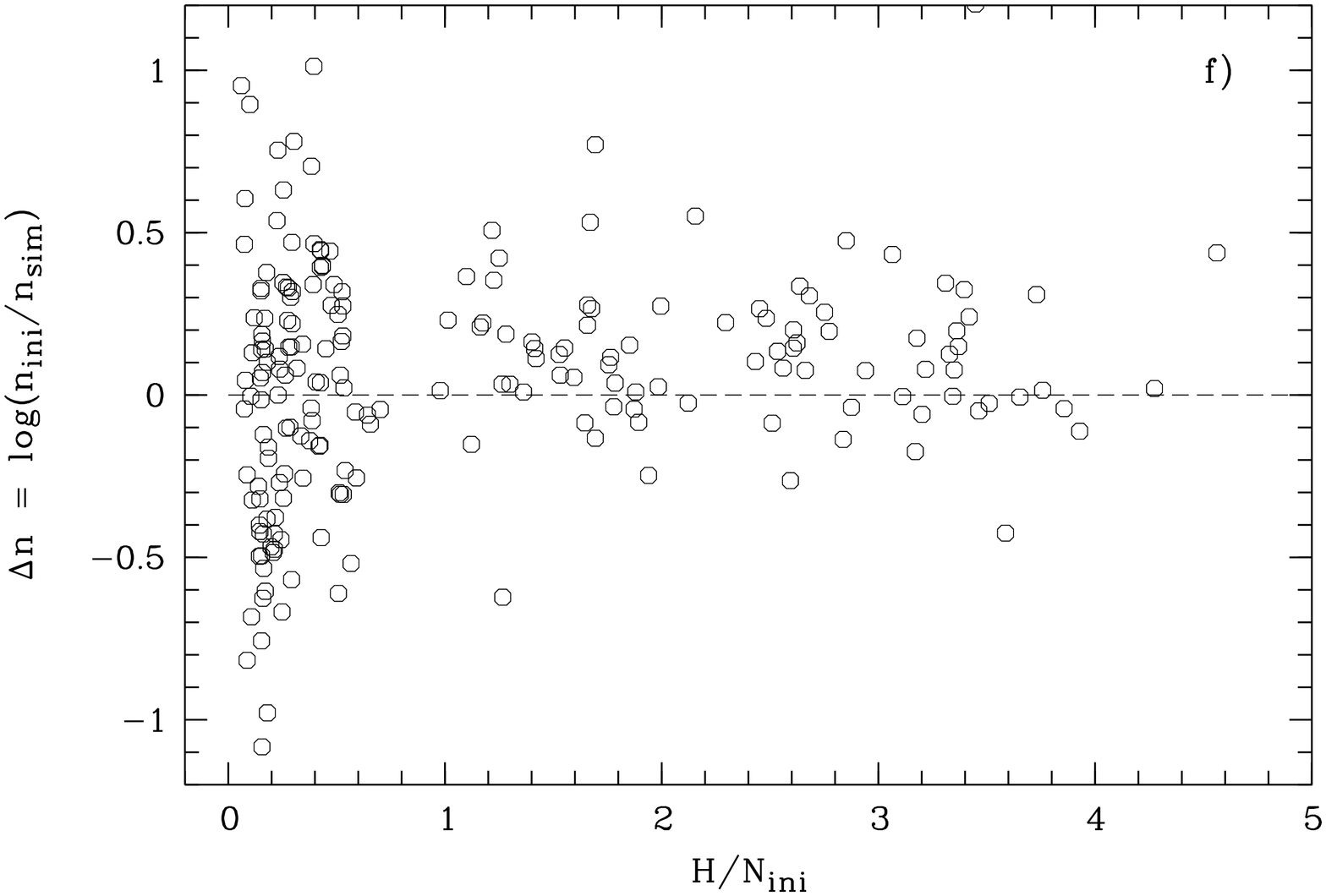}}
\caption{\label{galfitplots}
GALFIT recovery of host galaxy parameters for the simulated AGN sample. 
\emph{a)} Deviations between input host magnitude ${\rm v}_{\rm host,ini}$ 
and recovered host
  magnitude as a function of input host magnitude.
\emph{b)} Deviations between original host magnitude and recovered host
  magnitude ${\rm v}_{\rm host,sim}$
as a function of input $H/N_{\rm ini}$.
\emph{c)} Deviations between input effective radius ${\rm r}_{\rm eff,ini}$ 
and recovered effective radius ${\rm r}_{\rm eff,sim}$
as a function of input effective radius.
\emph{d)} Deviations between input effective radius
and recovered effective radius
as a function of input  $H/N$.
\emph{e)} Deviations between input S\'ersic index ${\rm n}_{\rm ini}$ 
and recovered S\'ersic index ${\rm n}_{\rm sim}$
as a function of input host magnitude.
\emph{f)} Deviations between input S\'ersic index ${\rm n}_{\rm ini}$ 
and recovered S\'ersic index ${\rm n}_{\rm sim}$
as a function of input $H/N$.
}
\end{figure*}

\begin{figure*}
\resizebox{\hsize}{!}{\includegraphics[height=8.5cm,angle=270]{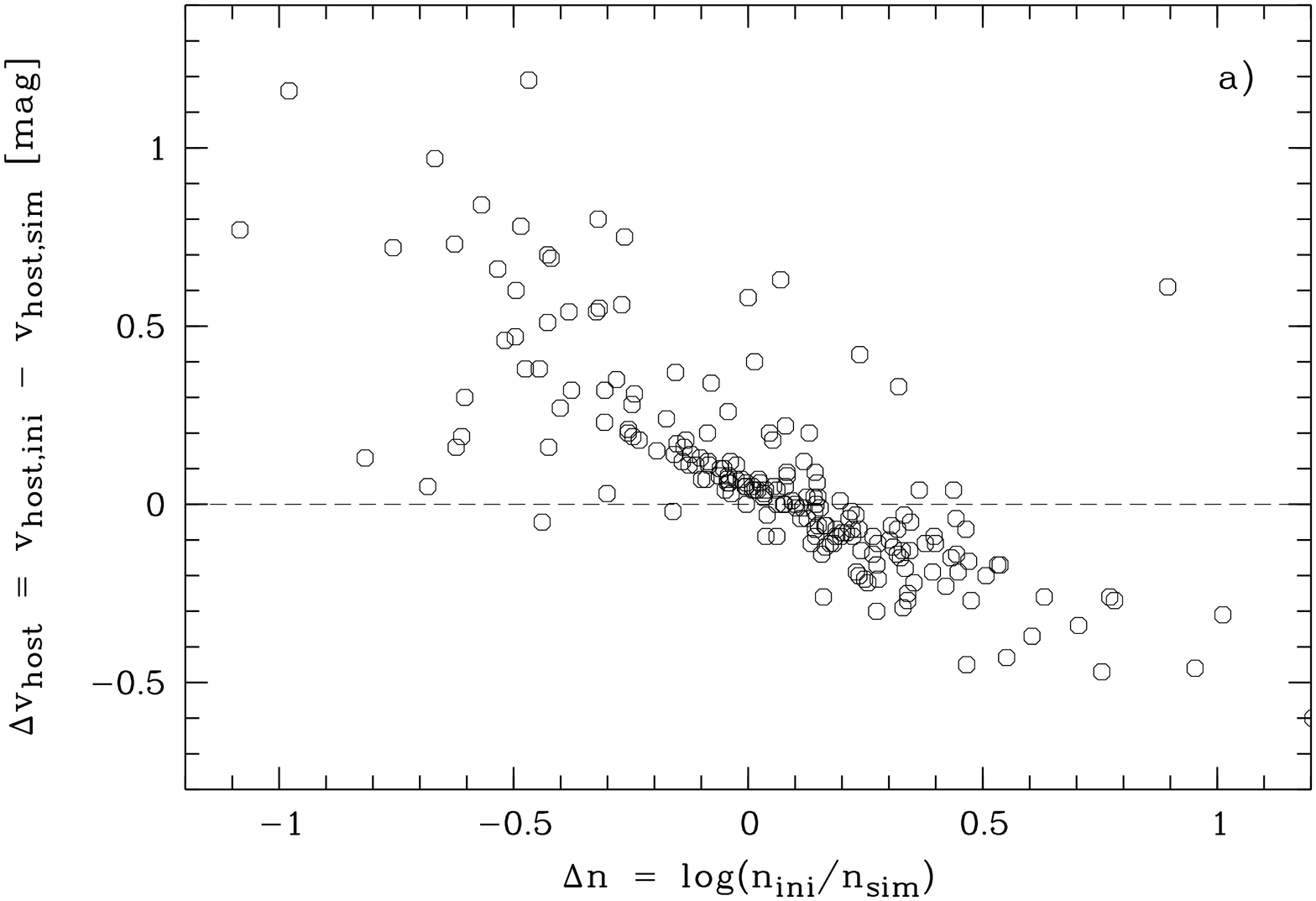}
\includegraphics[height=8.5cm,angle=270]{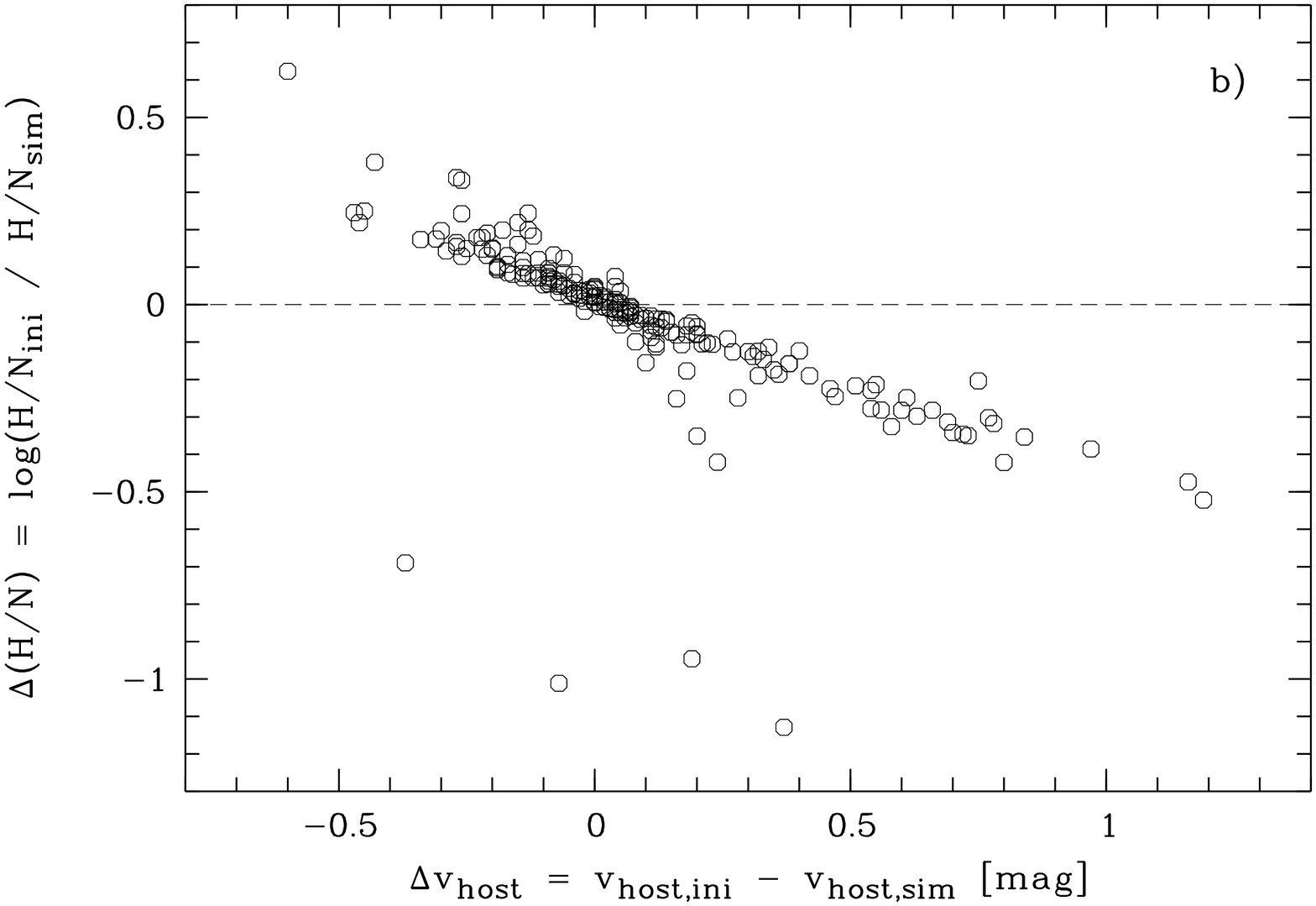}}
\resizebox{\hsize}{!}{\includegraphics[height=8.5cm,angle=270]{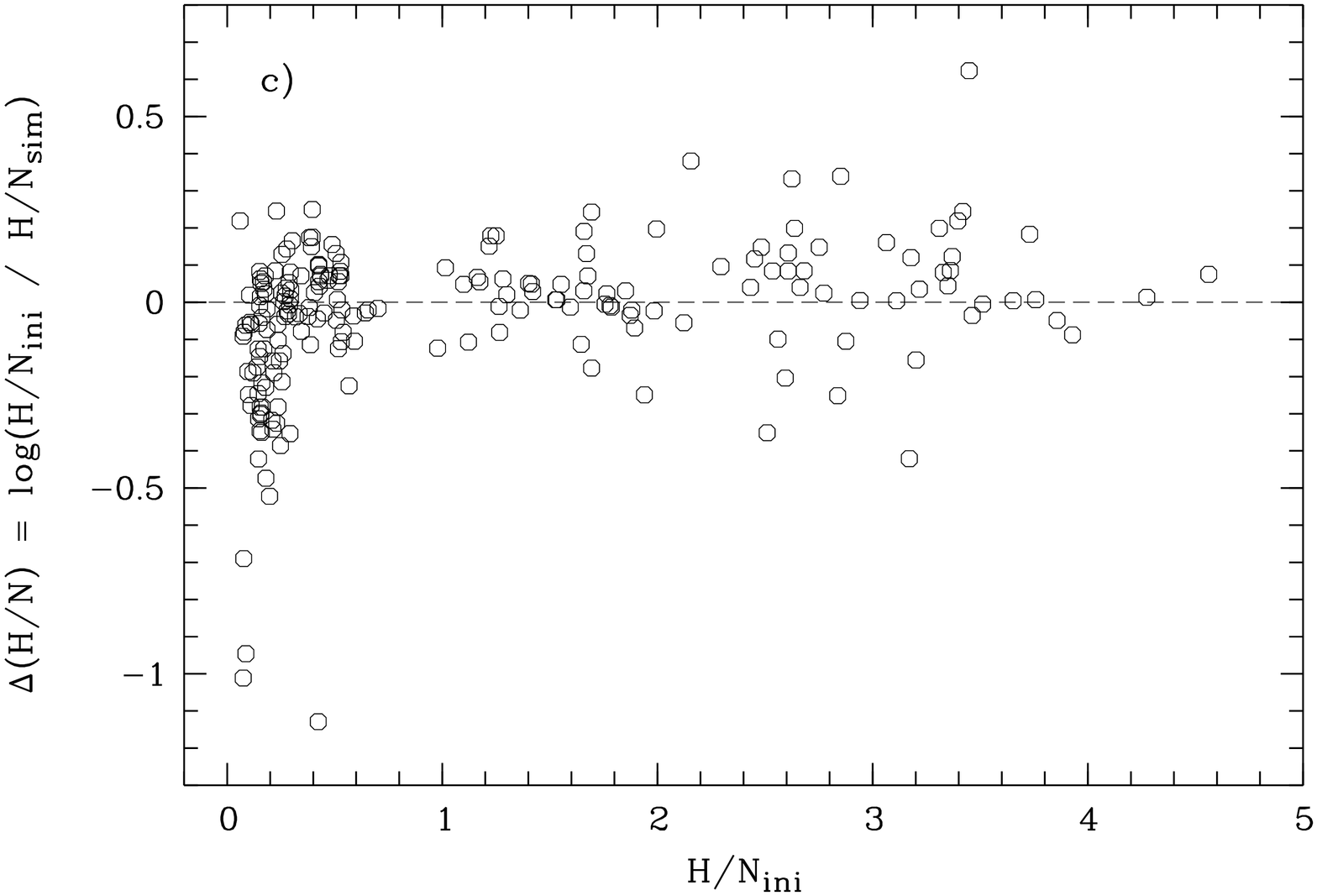}
\includegraphics[height=8.5cm,angle=270]{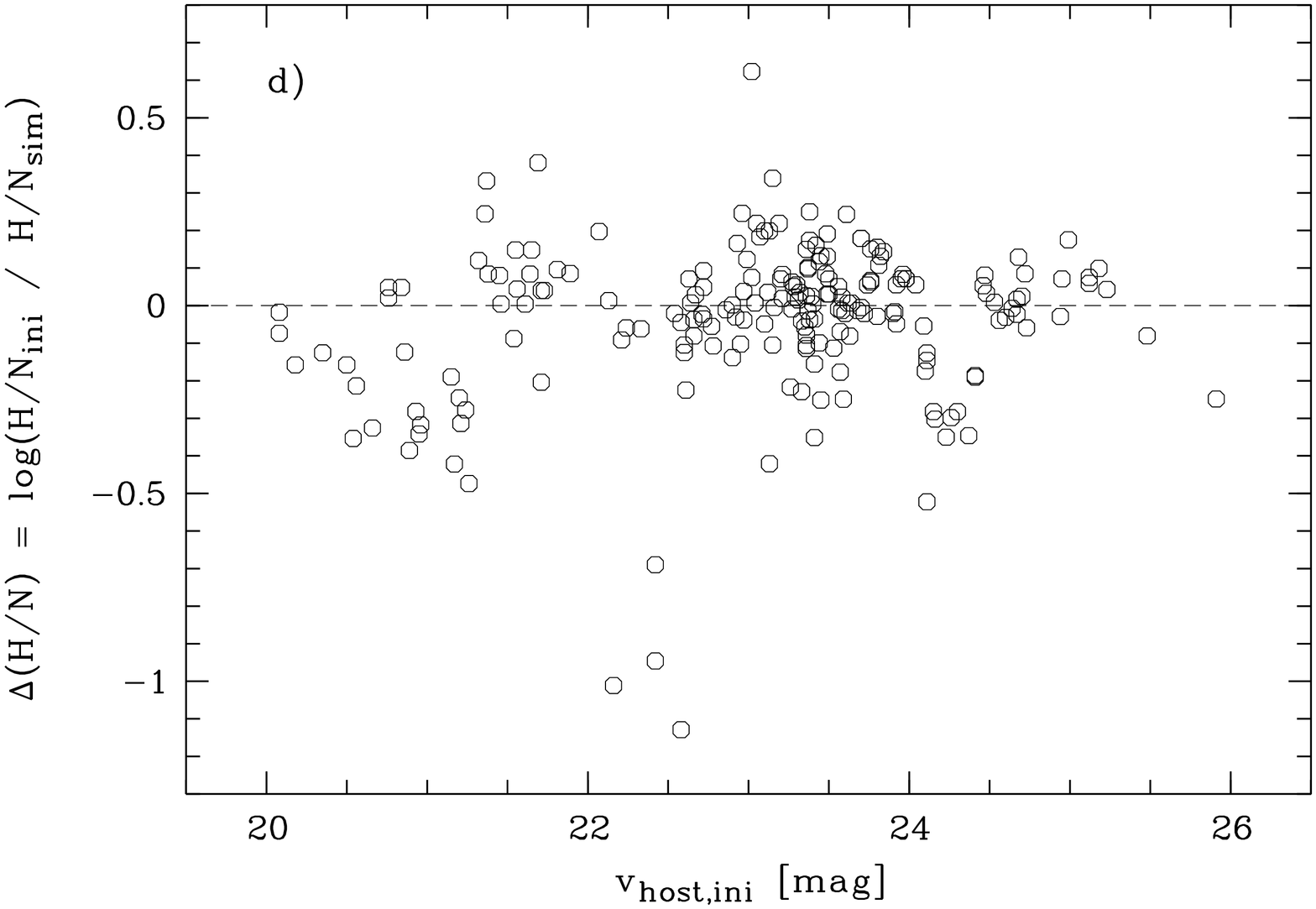}}
\caption{\label{galfitplots2}
GALFIT recovery of host galaxy parameters for the simulated AGN sample. 
\emph{a)} Deviations between input host magnitude ${\rm v}_{\rm host,ini}$ 
and recovered host magnitude ${\rm v}_{\rm host,sim}$ 
as a function of the deviations between input S\'ersic index ${\rm n}_{\rm ini}$ 
and recovered S\'ersic index ${\rm n}_{\rm sim}$.
\emph{b)} Deviations between input ${\rm H/N}_{\rm ini}$ 
and the recovered ${\rm H/N}_{\rm sim}$ 
as a function of the deviations between original host magnitude and recovered host
  magnitude.
\emph{c)} Deviations between input ${\rm H/N}_{\rm ini}$ 
and the recovered ${\rm H/N}_{\rm sim}$ 
as a function of input $H/N$.
\emph{d)} Deviations between input ${\rm H/N}_{\rm ini}$ 
and the recovered ${\rm H/N}_{\rm sim}$ 
as a function of input host magnitude.
}
\end{figure*}

Even though studying the robustness of
AGN host galaxy parameters derived with GALFIT is not the 
{\it raison d'\'etre} of
our analysis, it is an interesting byproduct.
The plots in Figs.~\ref{galfitplots} and
\ref{galfitplots2}
 compare the initial host parameters
(with suffix ini), 
as derived on the images \emph{prior}
to adding a synthetic nucleus, and the recovered host parameters 
(suffix sim)
from
simultaneous fitting of galaxy \emph{and} synthetic nucleus.

%
%
%

In Fig.~\ref{galfitplots}a, we show the total brightness $v_{\rm host,ini}$ of
the quiescent galaxies (single S\'ersic modeling on the original images,
\emph{before} the addition of synthetic nuclei) 
plotted against the deviations
$\Delta {\rm v}_{\rm host} = {\rm v}_{\rm host,ini}-{\rm v}_{\rm host,sim}$
between the initial brightness ${\rm v}_{\rm host,ini}$ and the brightness 
${\rm v}_{\rm host,sim}$ determined from
two-component GALFIT modeling after adding the synthetic nucleus.
The values of $\Delta {\rm v}_{\rm host}$ are basically distributed around zero with a
scatter of $\sim$\,0.25\,mag. The scatter  is mostly independent of the galaxy
magnitude except for objects brighter than ${\rm v}_{\rm host,ini}\approx21.3$.
It is surprising that most of these galaxies 
have high positive values of $\Delta {\rm v}_{\rm host}$, meaning that the
fluxes of these objects are overestimated when they are fitted simultaneously
with a synthetic nucleus. It is quite obvious that this occurs because the
fitting algorithm attributes part of the nucleus flux to the host, because
the S\'ersic indices of most of these galaxies are overestimated, too~---
this leads
to low values of the fraction ${\rm n}_{\rm ini}/{\rm n}_{\rm sim}$ 
between S\'ersic index in the original image
and the simulated AGN, depicted by filled green circles; see also panel e).

Fig.~\ref{galfitplots}b shows the deviations between original host magnitude
and recovered host  magnitude 
as a function of input $H/N_{\rm ini}$.
The scatter in $\Delta {\rm v}_{\rm host}$ strongly increases toward low
values of $H/N_{\rm ini}$, as expected.
Moreover, in the regime $H/N_{\rm ini}<1$, i.e.~in cases where the nucleus
outshines the host, the recovered host magnitudes ${\rm v}_{\rm host,sim}$ tend
to be too bright. This trend indicates that part of the bright nucleus flux
is attributed to the host component by the fitting algorithm.
A similar results has been found by Kim et al.~(2008). These authors argued
that mismatches between the nucleus profile and the input local PSF cause this
systematic effect.

Fig.~\ref{galfitplots}c shows the input effective radii ${\rm r}_{\rm eff,ini}$ 
determined on the original images plotted against the deviations 
between input effective radius ${\rm r}_{\rm eff,ini}$ 
and recovered effective radius ${\rm r}_{\rm eff,sim}$.
This parameter is accurate to within less than 30\% for 3/4 of the
objects.
Fig.~\ref{galfitplots}d shows the deviations between input 
and recovered effective radius as a function of input 
$H/N_{\rm ini}$. Like the host brightnesses,
the effective radii have a larger scatter toward low $H/N$.

In Fig.~\ref{galfitplots}e, the ratio 
$\Delta {\rm n} = \log({\rm n}_{\rm ini}/{\rm n}_{\rm sim})$ 
between original and recovered S\'ersic index is
plotted as a function of input total host magnitude 
${\rm v}_{\rm host,ini}$.
Evidently, n is a parameter that is much harder to
determine than the brightness or effective radius of the host galaxy. 
Toward fainter host magnitudes, 
the S\'ersic indices are systematically underestimated
in a decomposition into nucleus and host component. For the distribution in
$H/N$ considered here (the median for our AGN sample is 
$\langle H/N \rangle = 0.63$),
this effect starts to kick in at host magnitudes fainter than 
${\rm v}_{\rm host,ini}
\approx 23$.
Owing to this systematic effect and the very large scatter in $\Delta {\rm n}$, 
it is impossible to use the best-fit S\'ersic index to determine
the morphological type of an AGN host.  
A similar trend has been found e.g.~by 
Sanchez et al.~(\cite{san04}) on the basis of
simulated  
host galaxies and nuclei. Those simulations used symmetric
single-component host galaxies (S\'ersic profiles with either 
${\rm n}=1$ or ${\rm n}=4$).
In contrast to this, many of the host galaxies in our sample consist of two 
or more components (bulge, disk, potentially spiral structure, etc.).
This higher complexity of the host profiles leads to a much larger uncertainty 
in the recovered S\'ersic indices than found by Sanchez et al.
A comparison to the results of Simmons \& Urry~(\cite{sim08}) is difficult 
because the authors did not use constraints on the S\'ersic index in their fits.

Fig.~\ref{galfitplots}f displays the ratio 
$\Delta {\rm n} = \log({\rm n}_{\rm ini}/{\rm n}_{\rm sim})$ 
between original and recovered S\'ersic index as a function of the
input $H/N$.
At low $H/N$, the scatter in $\Delta {\rm n}$
is approximately 0.4\,dex.

Fig.~\ref{galfitplots2}a shows the host magnitude deviations
$\Delta {\rm v}_{\rm host}$
plotted as a function of the S\'ersic index deviations $\Delta {\rm n}$.
Not surprisingly, there is an anti-correlation of these two parameters.
A too faint model profile for the host galaxy ($\Delta {\rm v}_{\rm host}<0$)
is accompanied by an underestimated S\'ersic index, because a fraction of the
host galaxies' flux is assigned to the model nucleus. Since this effect
occurs at small galactocentric radii, the host model profile becomes too
shallow (vice versa for an overestimated host flux).
A false host model flux results in a wrong value of the
model $H/N$ (see Fig.~\ref{galfitplots2}b, 
displaying $\Delta H/N$ as a function of the host
magnitude deviations $\Delta {\rm v}_{\rm host}$).
Fig.~\ref{galfitplots2}c shows the ratio between the input 
$H/N_{\rm ini}$ and the recovered value $H/N_{\rm sim}$,
plotted as a function of $H/N_{\rm ini}$.
The $H/N$ is slightly underestimated for low input values 
and slightly overestimated for high input values, which partly 
is a consequence of the systematic error on ${\rm v}_{\rm host}$ at low
$H/N$.
In Fig.~\ref{galfitplots2}d, the ratio between the input and recovered $H/N$
is plotted against the original total host magnitude $v_{\rm host,ini}$.
In general, $H/N$ is overestimated for very bright hosts~--- which 
is induced by the frequently overestimated host fluxes in this regime
(cf. Fig.~\ref{galfitplots}a).

The comparisons between single-component light profile modeling of quiescent
galaxies and two-component modeling of these galaxies with 
added synthetic nuclei
provides a systematic error in the decomposition of AGN images using
GALFIT. Ignoring extreme outliers and taking only the best 95\% of the fits
into account, 
host fluxes, $H/N$, effective radii  and S\'ersic indices
are recovered to within fractional errors of 
$\sigma_{\rm vhost}$=0.23, 
$\sigma_{\rm H/N}$=0.34, and 
$\sigma_{\rm reff}$=0.50,
$\sigma_{\rm n}$=0.90, respectively. 
Since the scatter in host magnitude, effective radius and S\'ersic index 
is dominated by objects with low $H/N$ (cf. Fig~\ref{galfitplots}b, d and f), 
we repeated this test by rejecting objects with $H/N<0.4$.
We then find respective fractional errors of
$\sigma_{\rm vhost}$=0.18, 
$\sigma_{\rm H/N}$=0.46, and 
$\sigma_{\rm reff}$=0.22,
$\sigma_{\rm n}$=0.82. 
In particular for the effective radius, these values are more representative for an AGN decomposition with not too faint hosts. The numbers also show that
the S\'ersic indices of distant type-1 AGN hosts are very hard to recover.

\section{Morphological analysis and discussion \label{morphana}}

Prior to the following analysis, the best-fit model nuclei from the
GALFIT decomposition were subtracted from all synthetic and real AGN images.

We define a galaxy's outer contours using 
the segmentation maps from Source Extractor
because they determine which pixels of an image
are assigned to a given 
galaxy. All these maps were visually inspected. In a few
cases, e.g.~when the deblending of multiple galaxies was not satisfying, 
Source Extractor was re-run with different input parameters until the
segmentation maps were accurate. 
In our approach, major merger candidates were not deblended
into the individual 
sub-components but analyzed as one single object.
In the SExtractor configuration used here, the flux at
the outermost isophotes of a given galaxy corresponds to the 1\,$\sigma$ sky noise level.

\subsection{Definition of morphological descriptors}

We here rely on two parameter spaces
that have been frequently used in the literature, in particular for distant
galaxies. These are the concentration index C and asymmetry index A
(Conselice et al.~\cite{con00}), the Gini coefficient $G$ and the
\mtw\ index (Lotz et al.~\cite{lot04}).

The asymmetry parameter 
$A$ is measured by subtracting a galaxy image rotated by 180$^\circ$ from the
original image: 
\begin{equation}
A =  \min \bigg( \frac{\sum_{i,j} | I - I_{180}|}{\sum_{i,j} |I|} \bigg) - 
\min \bigg( \frac{\sum_{k,l} | B - B_{180}|}{\sum_{i,j}|I|}
\bigg).
\end{equation}
Here, $I$ is the original image, $I_{180}$ is the image rotated by
180$^\circ$ about
the galaxy's center, and $B$ and $B_{180}$ are the background and rotated
background. 
This latter contribution to $A$
 due to the background noise is estimated
from blank sky regions in the vicinity of the object under investigation.
The sums are computed over all pixels within the 1$\sigma$ isophotes 
of the galaxy; $B$ covers the same number of pixels as $I$.  
The minimization is computed by shifting the galaxy image on 
an integer pixel grid, with a maximum shift of eight pixels for our 
data sets.

There are various definitions of the concentration index $C$ in the literature that
only have slight differences to each other. 
We adopted $C$ as the ratio between two fluxes 
in elliptical areas $G_1$ and $G_2$, 
which have the same axis ratio and
position angle but a different major axis with $a_2 = 0.3\,a_1$ 
(Abraham et al.~\cite{abr96}):
\begin{equation}
\label{cdef} C = \frac{\sum_{i,j \in G_2} I_{ij}}{\sum_{k,l \in G_1} I_{kl}}.
\end{equation}
Here, $I_{ij}$ is the flux at the pixel $(i,j)$ within the inner isophotal region
$G_2$ and $I_{kl}$ is the flux at the pixel $(k,l)$ within the larger 
isophotal region $G_1$. 

The Gini coefficient has originally been developed for applications in social
sciences. When computed for galaxy images, it is a measure of the concentration
of an object's light profile, similar to $C$, but insensitive to the
definition of a central pixel.
We adopted the Gini coefficient as given in Lotz et al.~(\cite{lot04}):
\begin{equation}
G = \frac{1}{\bar{|f|} n (n-1)} \sum^n_i (2i - n -1) |f_i|,
\end{equation}
where $|f_i|$ are the absolute flux values of the $i$-th pixel (ranked in increasing
order) and $\bar{|f|}$ is the mean of the absolute flux values of all $n$ pixels
assigned to an object.

The \mtw\ index is based on the total second-order moment $M_{tot}$, which  
is the flux in each pixel $f_i$ multiplied by the squared distance to the
center of the galaxy, 
summed over all pixels of a given object:
\begin{equation} 
M_{tot} = \sum_i^n M_i = \sum_i^n f_i \cdot ((x_i - x_c)^2 + (y_i - y_c)^2),
\end{equation}
where $x_c, y_c$ is the galaxy's center. The center is computed by finding $x_c, y_c$ 
such that $M_{tot}$ is minimized.
$M_{20}$ then is defined as the normalized second-order moment of the brightest 20\% of
the galaxy's flux.  To compute $M_{20}$, the pixels are rank-ordered by flux, and
$M_i$ is summed over the brightest pixels until the sum of the brightest
pixels amounts to 20\% of the 
total galaxy flux, followed by a normalization to $M_{tot}$ (Lotz et al.~\cite{lot04}):
\begin{eqnarray}
M_{20} = \log \left(\frac{\sum_i M_i}{M_{tot}}\right) & {\rm while } & \sum_i
f_i <  0.2 f_{tot},
\end{eqnarray}
Here, $f_{tot}$ is the total flux of a galaxy 
and $f_i$ are the fluxes for each pixel $i$, 
ordered according to their flux values such that $f_1$ is the brightest pixel,
$f_2$ the second brightest, etc. Basically, \mtw\ traces the spatial distribution 
of bright features such as off-center star-forming regions.

\subsection{Quantitative morphological comparison
 of AGN hosts and quiescent galaxies\label{finana}}

Our visual classifications of the comparison sample will be used as a
calibration in the $C/A$ and $G$/\mtw\ parameter spaces. Since these
classifications were carried out prior to including a synthetic
optical nucleus, they are much more robust than
the nucleus-subtracted host galaxy images of the real AGN that
are affected by the nucleus residuals in the center of
the images (cf. Fig.~\ref{galfitplots}).

First we defined when to call a visual classification reliable.
Eight team members visually inspected the galaxies in the non-AGN
comparison sample and classified them as a major merger, undisturbed, or
unclear. 
Combining the various classifications for each object, we constructed
various samples (as described in Sect.~\ref{vis}).
Only using cases with at least four identical classifications, we obtained a
sample 4/8, five or more identical classifications define a 
sample 5/8, etc. 
We used a two-dimensional Kolmogorov-Smirnov test to quantify
the distinguishability between major merger candidates and undisturbed
galaxies in the parameter spaces $C$/$A$ and $G$/\mtw.
Table~\ref{kstest1} shows the respective probabilities $P$ that both samples
are drawn from the same parent distribution. The probability $P$ increases
with the minimum number of agreeing classifications.
This mainly is an effect of the sample sizes. As a test, we restricted
all samples to the size of the 7/8 scheme (by randomly selecting subsets
from the other larger samples) and recomputed the KS test. 
The probabilities then are comparable between the four different
schemes. 
However, since a larger sample offers more robust statistics, we decided to
use scheme 4/8 in the following.

Our main morphological analysis is based on galaxy images 
with removed central point sources (cf.~Fig.~\ref{simul}).
For this reason we tested the distinguishability between major mergers
and undisturbed galaxies also with the images of the
\emph{nucleus-subtracted simulated AGN}.
As shown in Table~\ref{kstest2}, the contrast between the two 
morphologies is degraded with respect to the original images,
because of the residuals in the galaxy centers after the point-source
subtraction (just for completeness, we give the KS test results for all
four classification schemes).

\begin{table}[t] 
\caption{\label{kstest1}Two-dimensional KS test comparing major mergers 
and undisturbed galaxies in the quiescent comparison sample.}
\begin{center}
\begin{tabular}{ccc} \hline\hline
 & &  \\
Sample & $C$/$A$ &  $G$/\mtw  \\ 
(1) & (2) & (3)  \\
 & &  \\ \hline
 & &  \\
4/8 & $10^{-6}$  & $8 \times 10^{-6}$ \\
5/8 & $2 \times 10^{-6}$   & $1.8 \times 10^{-5}$  \\
6/8 & $4 \times 10^{-5}$  & $3.1 \times 10^{-5}$  \\
7/8 & $1.1 \times  10^{-3}$  & $9.1 \times 10^{-4}$ \\ 
 & &   \\ \hline
\end{tabular}
\end{center}
\emph{Note:} Col.~(1) denotes the required minimum number of classifiers agreeing 
on the type of a given object;
Col.~(2): probability $P$ that both samples stem from the same parent 
distribution in $C$/$A$ space; Col.~(3): same for $G$/\mtw\ space.
\end{table}

\begin{table}[t] 
\caption{\label{kstest2}Two-dimensional KS test comparing major mergers 
and undisturbed galaxies in the simulated AGN sample (after removal of the
central point sources).}
\begin{center}
\begin{tabular}{ccc} \hline\hline
 & &  \\
Sample & $C$/$A$ &  $G$/\mtw  \\ 
(1) & (2) & (3)  \\
 & &  \\ \hline
 & &  \\
4/8 & $8.6 \times 10^{-4}$  & $1.1 \times 10^{-3}$ \\
5/8 & $5.7 \times 10^{-4}$   & $2.7 \times 10^{-3}$  \\
6/8 & $5.2 \times 10^{-3}$  & $3.9 \times 10^{-2}$  \\
7/8 & $1.1 \times  10^{-2}$  & $4.4 \times 10^{-3}$ \\ 
 & &   \\ \hline
\end{tabular}
\end{center}
\emph{Note:} Columns are the same as in Table~\ref{kstest1}.
\end{table}

\begin{figure*}
\resizebox{\hsize}{!}{\includegraphics[height=9cm,angle=270]{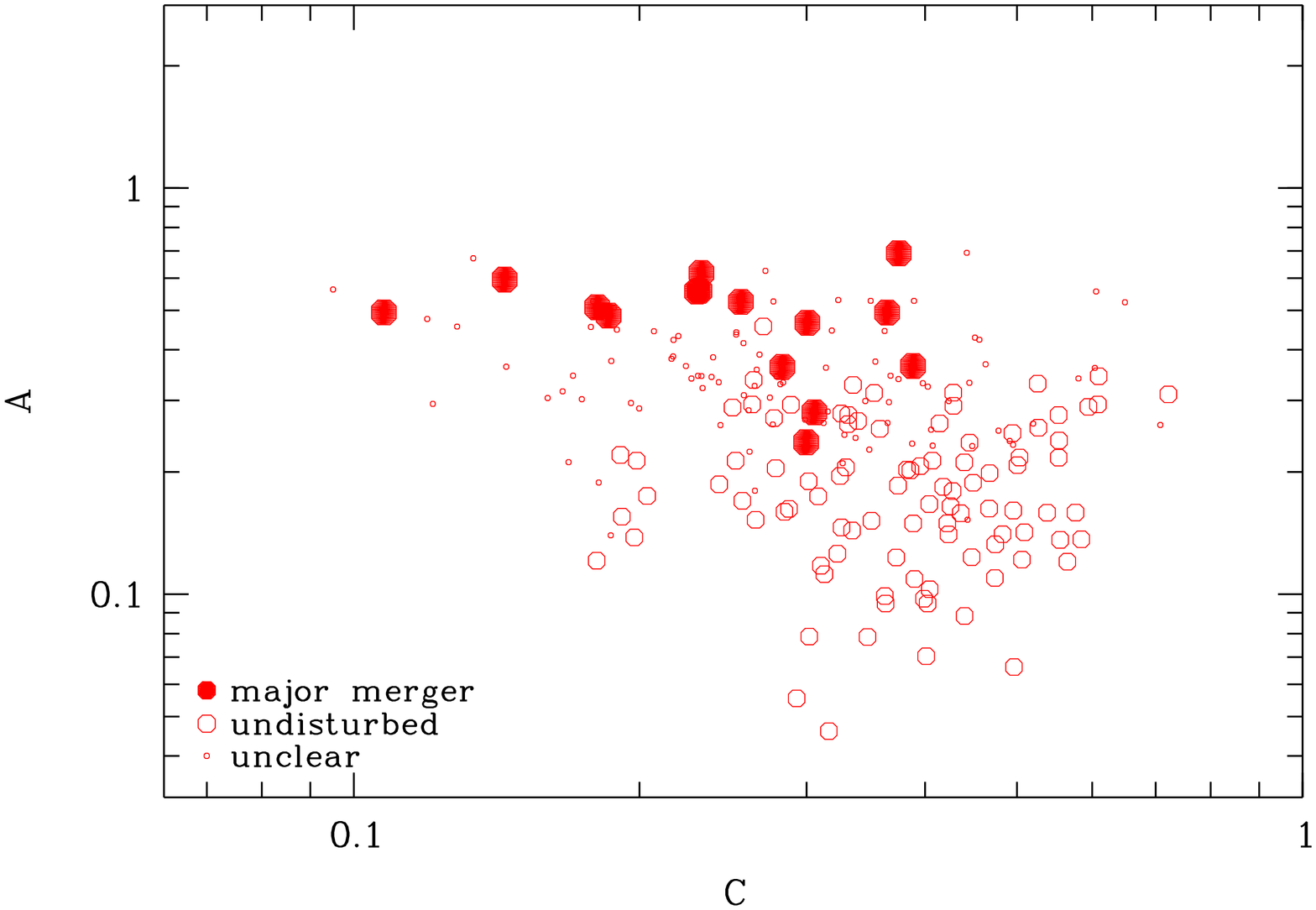}
\includegraphics[height=9cm,angle=270]{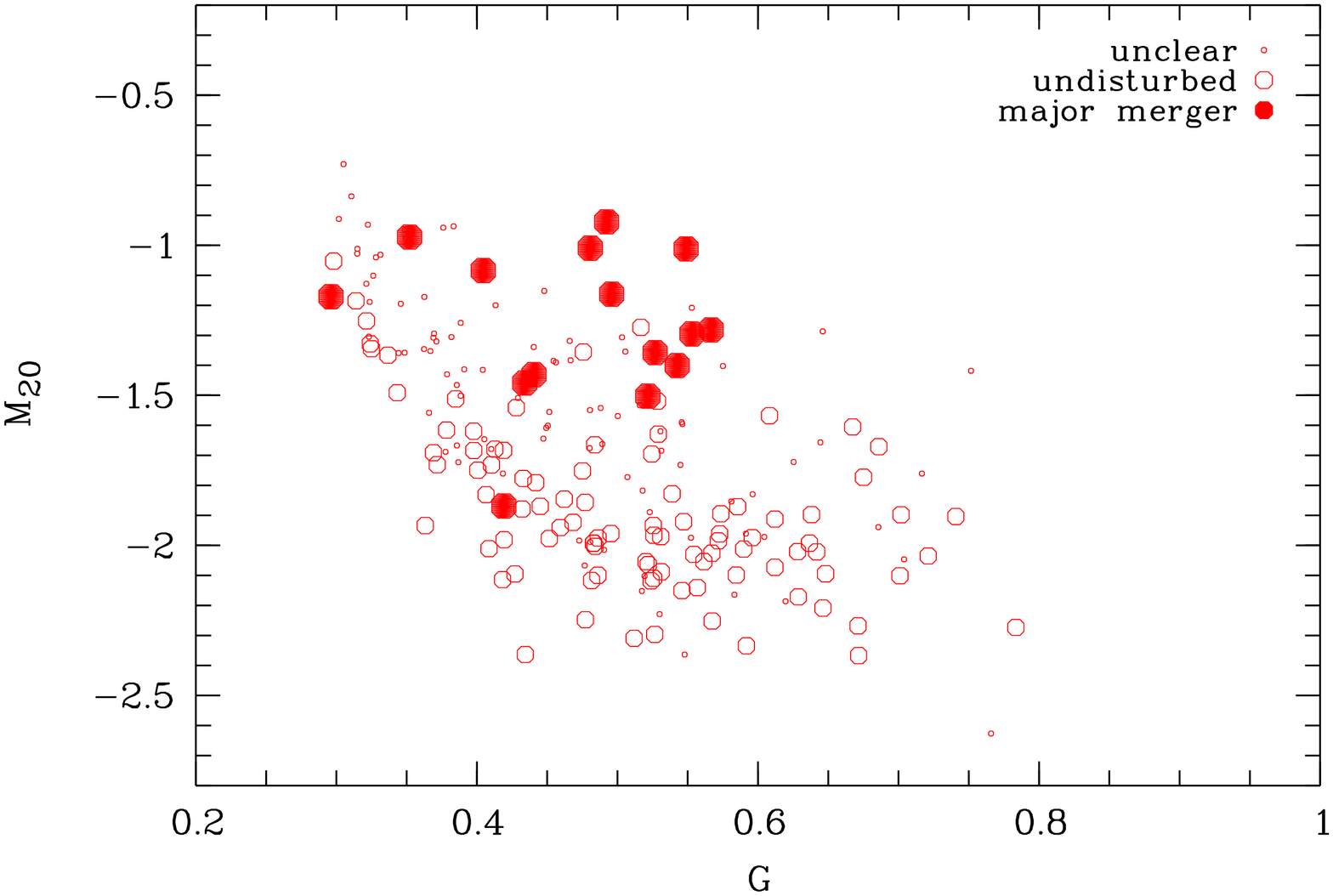}}
\caption{\label{orig}
Distribution of the quiescent galaxies in concentration
index $C$ versus asymmetry $A$ (left) and in
Gini coefficient $G$ versus \mtw\ (right).
Tiny open symbols depict ambiguous cases, large open
symbols show undisturbed galaxies, large filled symbols are 
classified as major mergers. Galaxies have highly
concentrated light profiles toward the right-hand side of the graphs and
become more asymmetric / show more substructure toward the upper part
of the graphs.}
\end{figure*}

In Fig.~\ref{orig} (left-hand plot), 
we show the distribution of the quiescent galaxies in
terms of $C$ vs. $A$. Major merger candidates have systematically
higher asymmetry indices $A$ and lower concentration $C$ than undisturbed
galaxies. This is well-known from observations as well as from
synthetic images created from
numerical simulations (e.g.~Lotz et al.~2008).
The galaxies with unclear morphology are located between major mergers
and undisturbed galaxies, showing substantial overlap with the former.
A number of galaxies that could not be clearly characterized visually
might thus be major mergers. Still, the KS test probability that
major mergers and unclear objects stem from the same parent distribution
is low: $P=0.004$.

%
%

%
%
\begin{figure*}
\resizebox{\hsize}{!}{\includegraphics[height=9cm,angle=270]{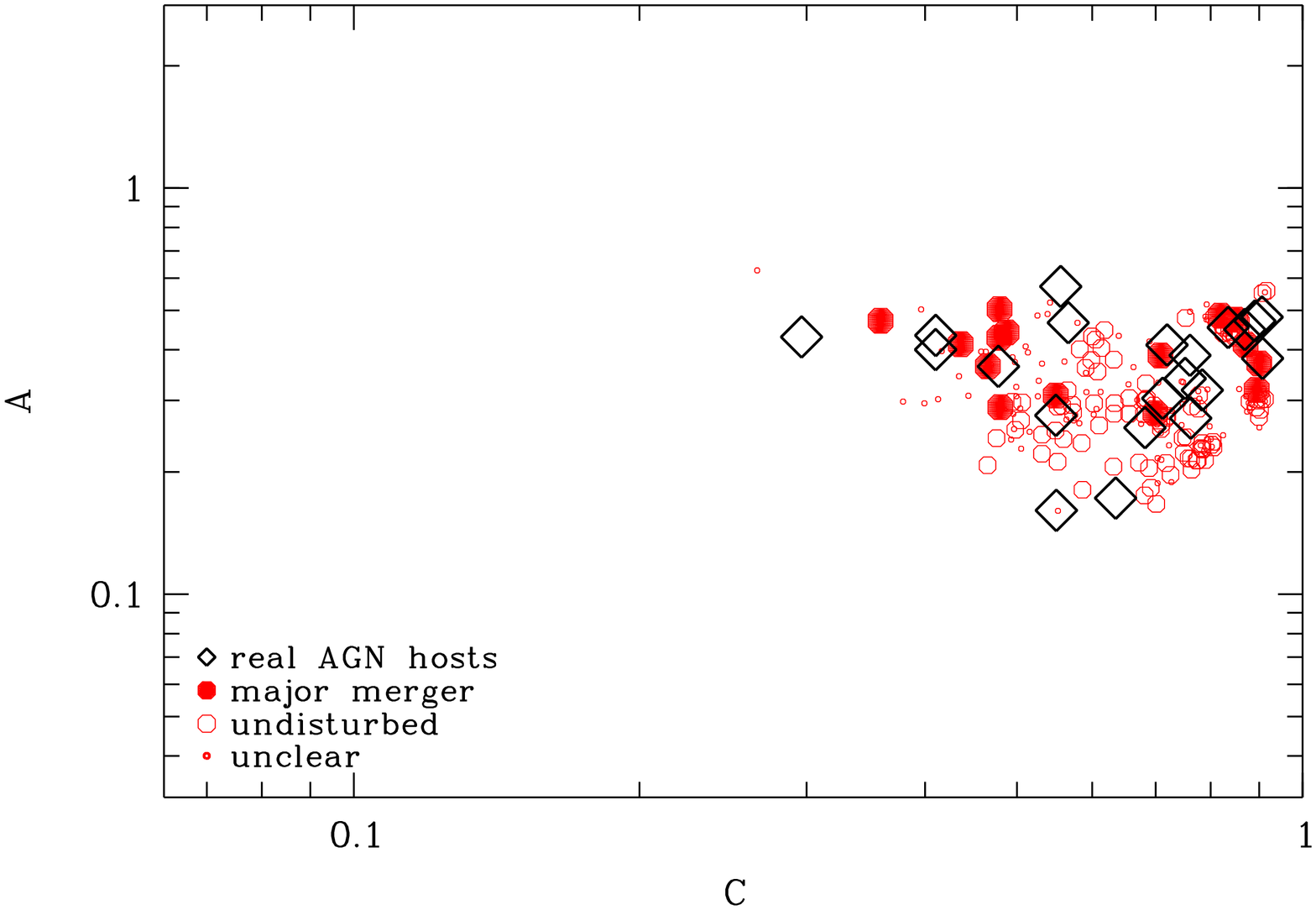}
\includegraphics[height=9cm,angle=270]{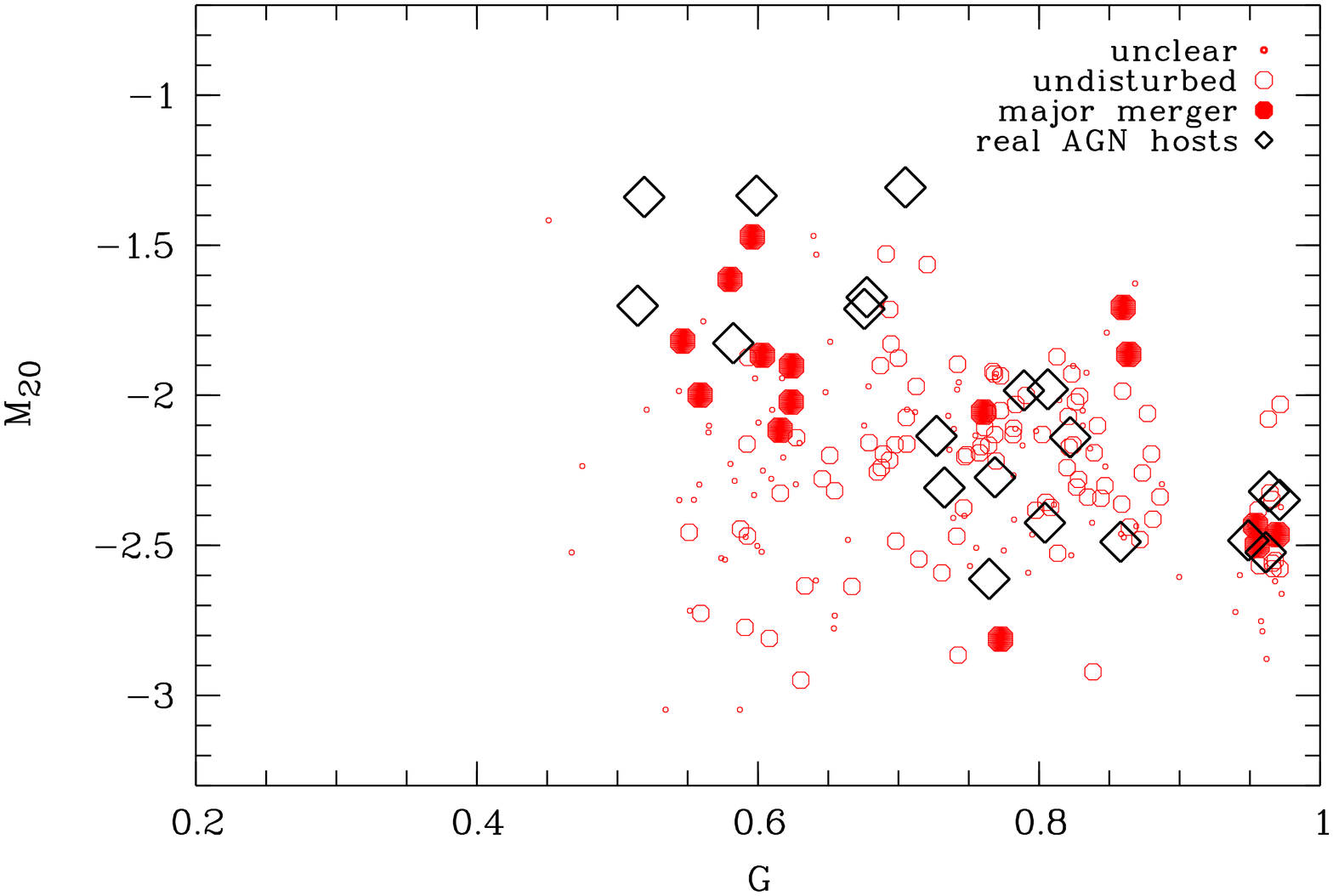}}
\caption{\label{unsub}
Distribution of the simulated AGN
(i.e. a quiescent galaxy with an additional synthetic nucleus
(\emph{before subtraction}), depicted here 
as circles) 
and real AGN host galaxies (open diamonds), in concentration
index $C$ versus asymmetry $A$ (left) and in
Gini coefficient $G$ versus \mtw\ (right).
Tiny open symbols depict ambiguous cases, large open
symbols show undisturbed galaxies, large filled symbols are 
classified as major mergers.
With respect to the original galaxy images (the distributions of which are
shown in Fig.~\ref{orig}), the distinguishability between different morphological types is degraded.
}
\end{figure*}

In $G$/\mtw\ space (right-hand plot of Fig.~\ref{orig}), 
the situation is similar: major mergers populate a region of high
\mtw\ values (i.e. bright off-center features) and lower $G$ indices
compared to the undisturbed galaxies. The distinction between the two classes
seems slightly worse than in $C$/$A$ space; this indeed is reflected in the
higher KS test $P$ values for $G$/\mtw\ with respect to $C$/$A$ 
(Table~\ref{kstest1}).

The distributions in $C/A$ and $G$/\mtw\
are completely different in the presence of an optical
nucleus. In  Fig.~\ref{unsub}, we show 
simulated AGN (quiescent galaxies plus synthetic nucleus)
and real AGN (\emph{including} their optical nuclei). 
Note that the displayed range in \mtw\ is slightly different
from Fig.~\ref{orig}.
Compared to the original quiescent galaxies, the synthetic AGN
have much higher concentration indices and
Gini coefficients, which is an obvious  effect of the
nuclear point source.
The \mtw\ indices become lower, indicating the loss of structure
because the combined light profile of host$+$nucleus is smoother than that of
the host alone.
The mean asymmetry of all types 
(median value $\langle A \rangle_{\rm sim}= 0.3$)
is almost unchanged with respect to the original galaxy
images ($\langle A \rangle_{\rm orig}= 0.27$). This is somewhat surprising
because the nuclear point sources should result in a less
elliptical combined light profile. 
However, 
the synthetic nucleus is usually not perfectly centered
on the pixel grid;
just like the nucleus in an observed real AGN. 
Since the asymmetry index is minimized on an 
integer pixel shift basis, even an almost symmetric light profile can 
therefore yield an
untypically high value for $A$. 
One other net effect of the nuclei was that the
distribution in $A$ became narrower.
The distinguishability between the morphological classes of the
simulated AGN hosts is, when compared to the initial distributions in 
Fig.~\ref{orig}, clearly degraded. 
Repeating the KS test to quantify how well undisturbed and major mergers 
are separated, we find $P=0.001$ in $C$/$A$ and 
$P=0.004$ in $G$/\mtw.
Simulated and real AGN cover similar regions in the parameter spaces.
Before we perform a quantitative comparison, we go one step further.

\begin{figure*}
\resizebox{\hsize}{!}{\includegraphics[height=9cm,angle=270]{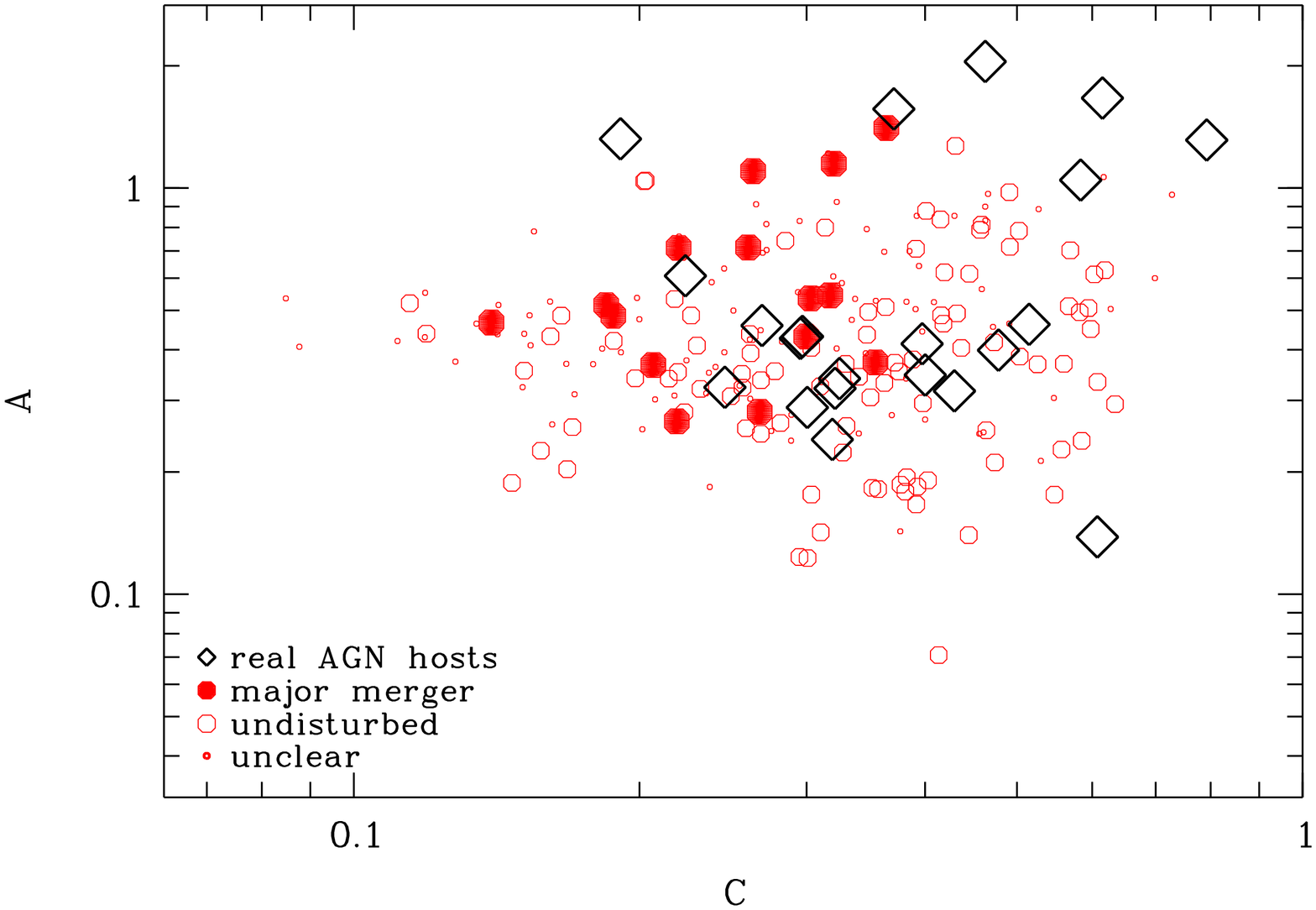}
\includegraphics[height=9cm,angle=270]{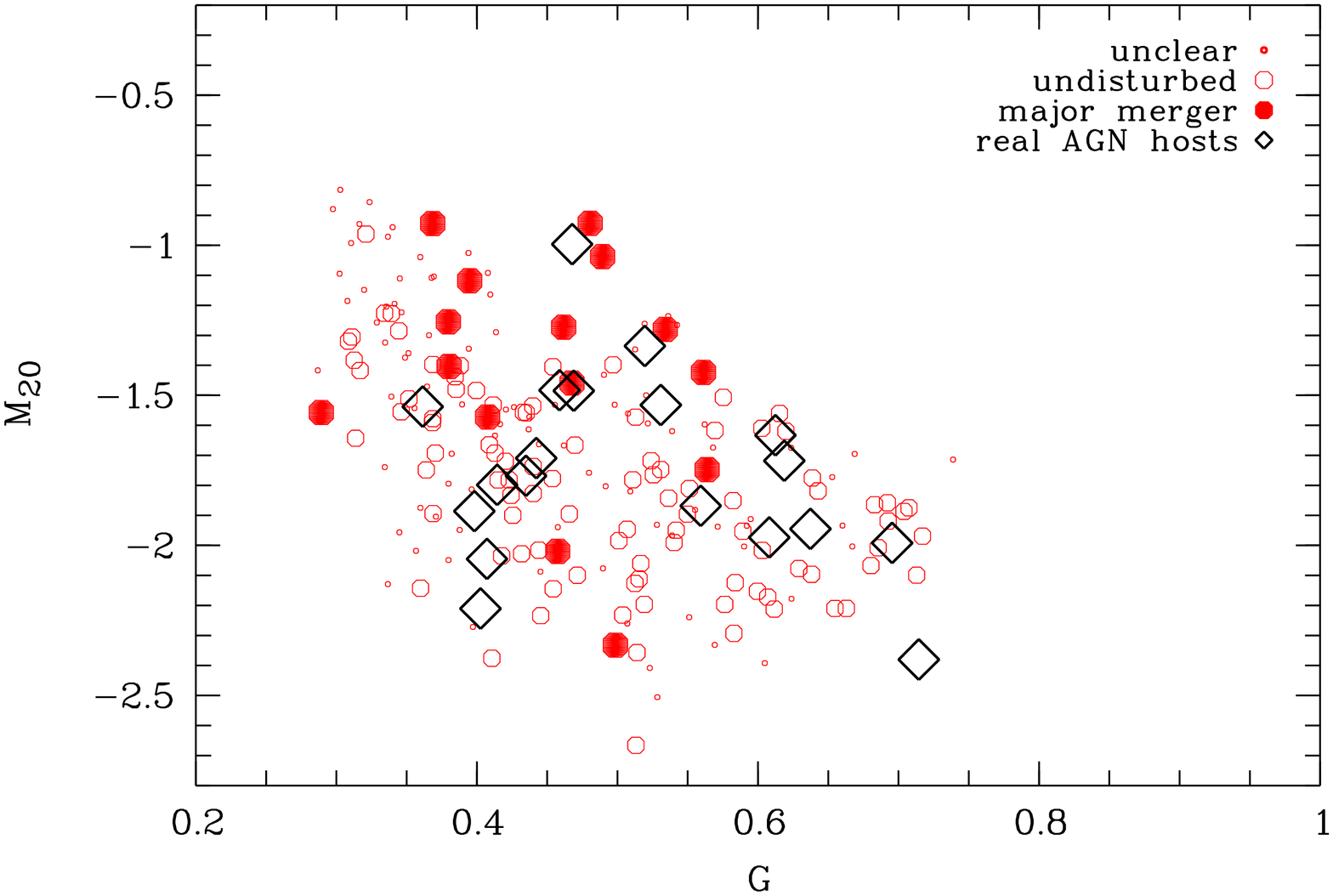}}
\caption{\label{subtracted2}
Distribution of the simulated and real AGN host galaxies after 
\emph{subtracting} the best-fit model nucleus in concentration
index $C$ versus asymmetry $A$ (left) and in
Gini coefficient $G$ versus \mtw\ (right).
Tiny open symbols depict ambiguous cases, large open
symbols show undisturbed galaxies, large filled symbols are 
classified as major mergers.
Major mergers and undisturbed galaxies are better separated than 
before subtracting the nuclei (Fig.~\ref{unsub}).
}
\end{figure*}

For the next computation of $C$, $A$, $G$, and \mtw, the best-fit nuclei
from two-component modeling with GALFIT were \emph{subtracted} 
from the simulated
and real AGN. The resulting distributions are shown in Fig.~\ref{subtracted2}.
In $C$/$A$ space, mergers and undisturbed objects
are better separated than before subtracting the nucleus
(cf.~Fig.~\ref{unsub}). However, there clearly is a stronger
overlap between these two types than there was based on the original images 
(cf.~Fig.~\ref{orig}).
This is reflected in an increase of the KS test probability when comparing 
merging and undisturbed galaxies: values increase
from $P=10^{-6}$  to $P \approx 10^{-3}$ in $C$/$A$ and
from $P \approx 10^{-5}$ to $P \approx 10^{-3}$ in $G$/\mtw,
as given in Tables \ref{kstest1} and
\ref{kstest2}.
Some objects feature asymmetry indices above unity, which in the case of
normal galaxy images would be impossible. This effect indicates 
additional noise in the center of the images that is not related to the sky 
background but residuals from the nucleus subtraction. 
As we discuss below, this particularly affects
objects with low $H/N$.

%
%

We will now use the KS test to determine which subsample of the quiescent
galaxies is most similar to the AGN hosts.
Or, more precisely, which subsample of the nucleus-subtracted 
\emph{synthetic} AGN hosts is most similar to the \emph{real} AGN hosts.
Comparing the AGN to major mergers, undisturbed, and unclear objects
in $C/A$ space, we find
$P_{\rm mm} = 0.015$, $P_{\rm undis} = 0.32$ and 
$P_{\rm uncl} = 0.034$, respectively.
The distribution of the AGN hosts in asymmetry and concentration
is by far
best matched by the undisturbed galaxies, while unclear cases
show the second highest~--- albeit much lower~--- probability.
The same comparison in $G$/\mtw\ space yields
$P_{\rm mm} = 0.040$, $P_{\rm undis} = 0.22$ and 
$P_{\rm uncl} = 0.033$, respectively.
Again, the distribution of undisturbed galaxies is most similar to the AGN
hosts, while 
major mergers are much less likely to stem from the same
parent distribution. 

We also compared the AGN hosts to the \emph{whole} quiescent
sample. The KS test then yields
$P_{\rm all}=0.20$ in $C/A$ or $P_{\rm all}=0.12$ in $G$/\mtw.
In both parameter spaces, these probabilities are slightly
lower than those 
comparing between AGN hosts and undisturbed galaxies.
This indicates that the AGN hosts comprise a  mix of
morphological classes that is slightly different from the 
comparison sample, in the sense that the former
contain a higher fraction of undisturbed galaxies than the latter.

These numbers argue against a significant 
fraction of major mergers among the AGN hosts.
Frequent minor mergers, on the other hand, cannot be ruled out. 
The unclear category is a mixed bag, containing minor mergers as well as
ambiguous cases that might partly be major mergers.
If major mergers are not the main triggering mechanism, these
ambiguous cases will result in a relatively low KS test probability,
just like we have found. 
We tested 
an alternative visual classification scheme that distinguishes between
minor and major mergers. To do this on an object-by-object basis
is of course a challenging task. 
The usage of this alternate scheme 
confirmed our results, albeit at lower statistical significance,
on the basis of KS statistics. 
This is why we here adopted a simple 
scheme with only three different morphological types, as
a straightforward test of the major merger scenario.

%
%
\begin{figure*}
\resizebox{\hsize}{!}{\includegraphics[height=9cm,angle=270]{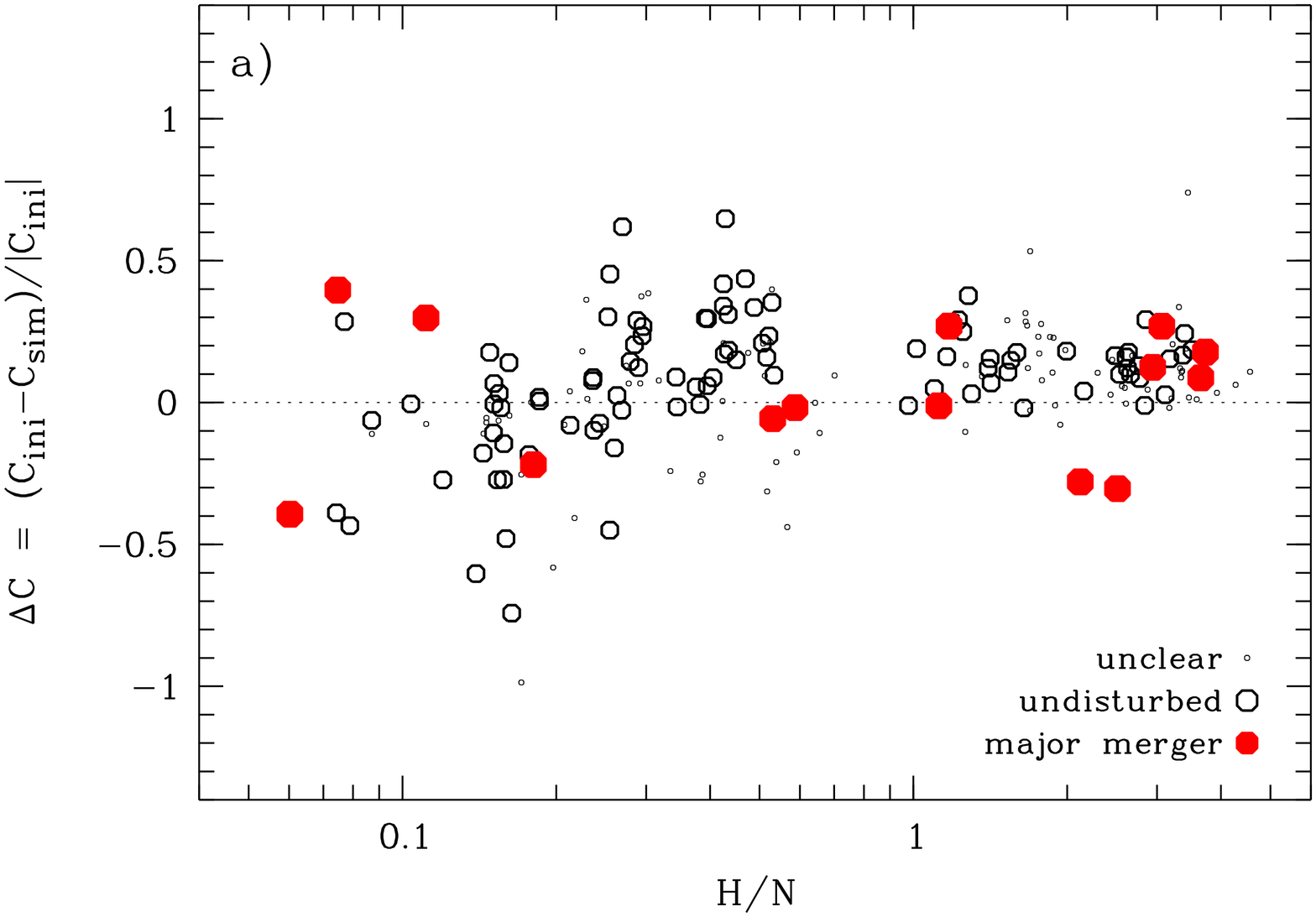}
\includegraphics[height=9cm,angle=270]{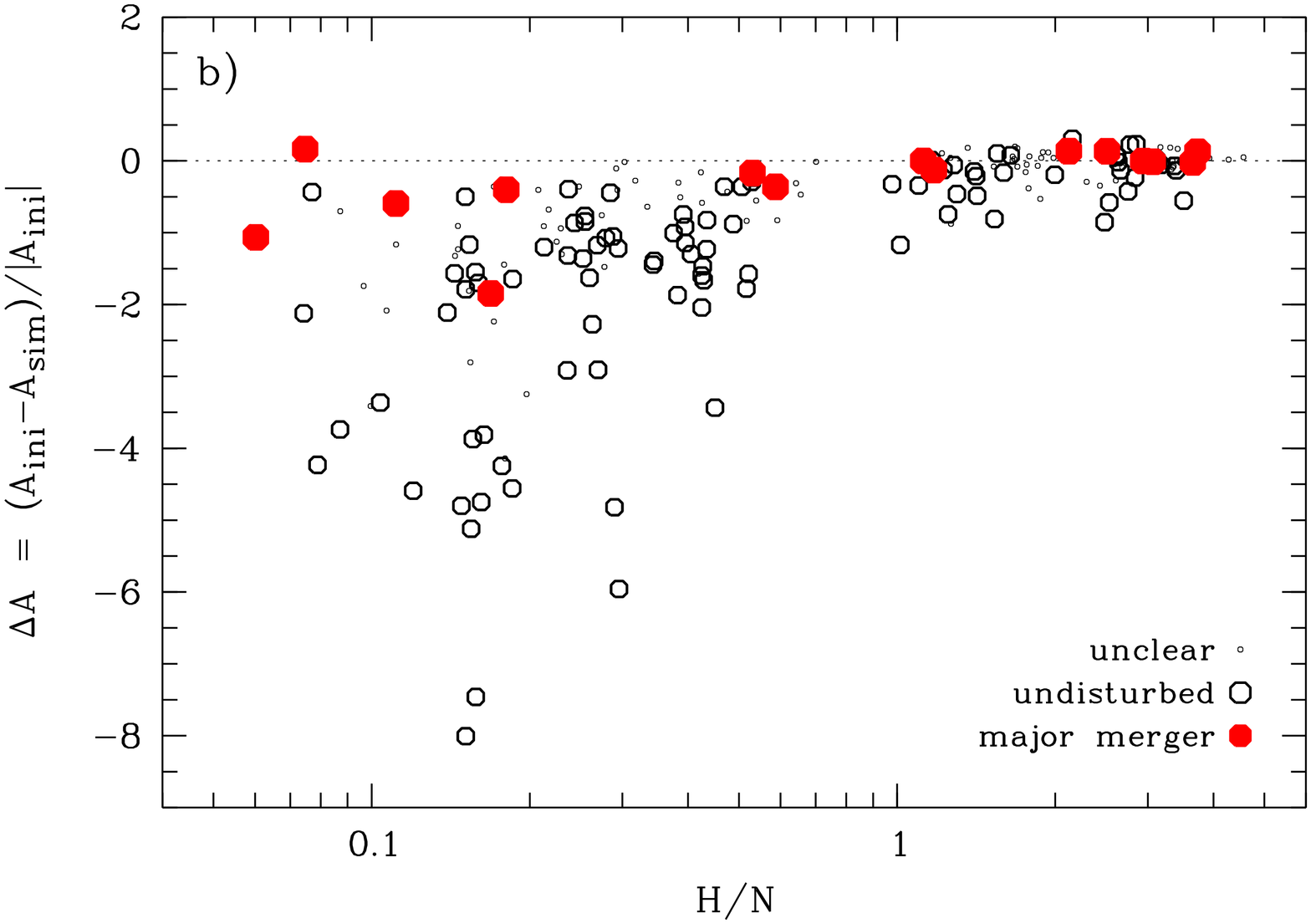}}
\resizebox{\hsize}{!}{\includegraphics[height=9cm,angle=270]{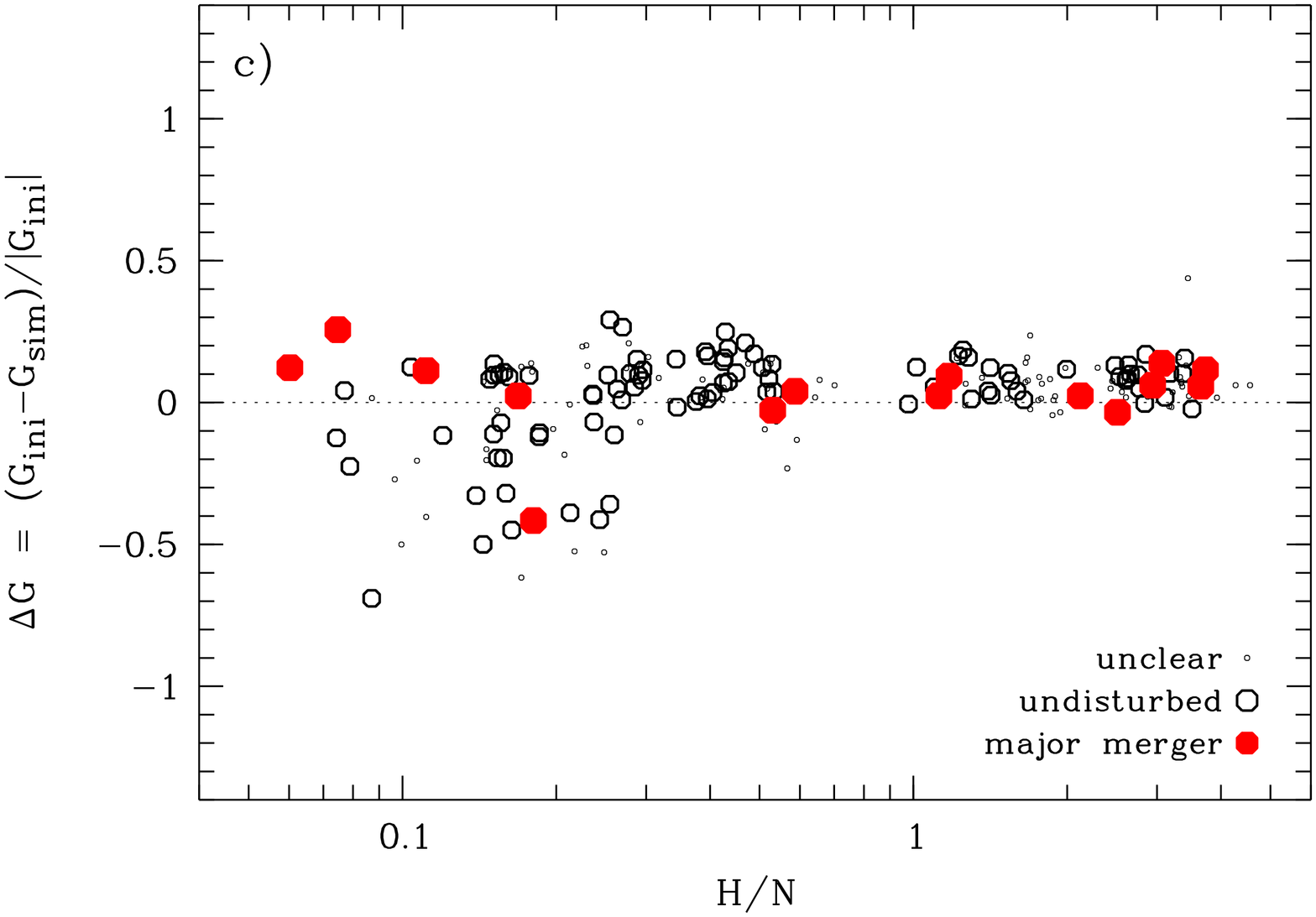}
\includegraphics[height=9cm,angle=270]{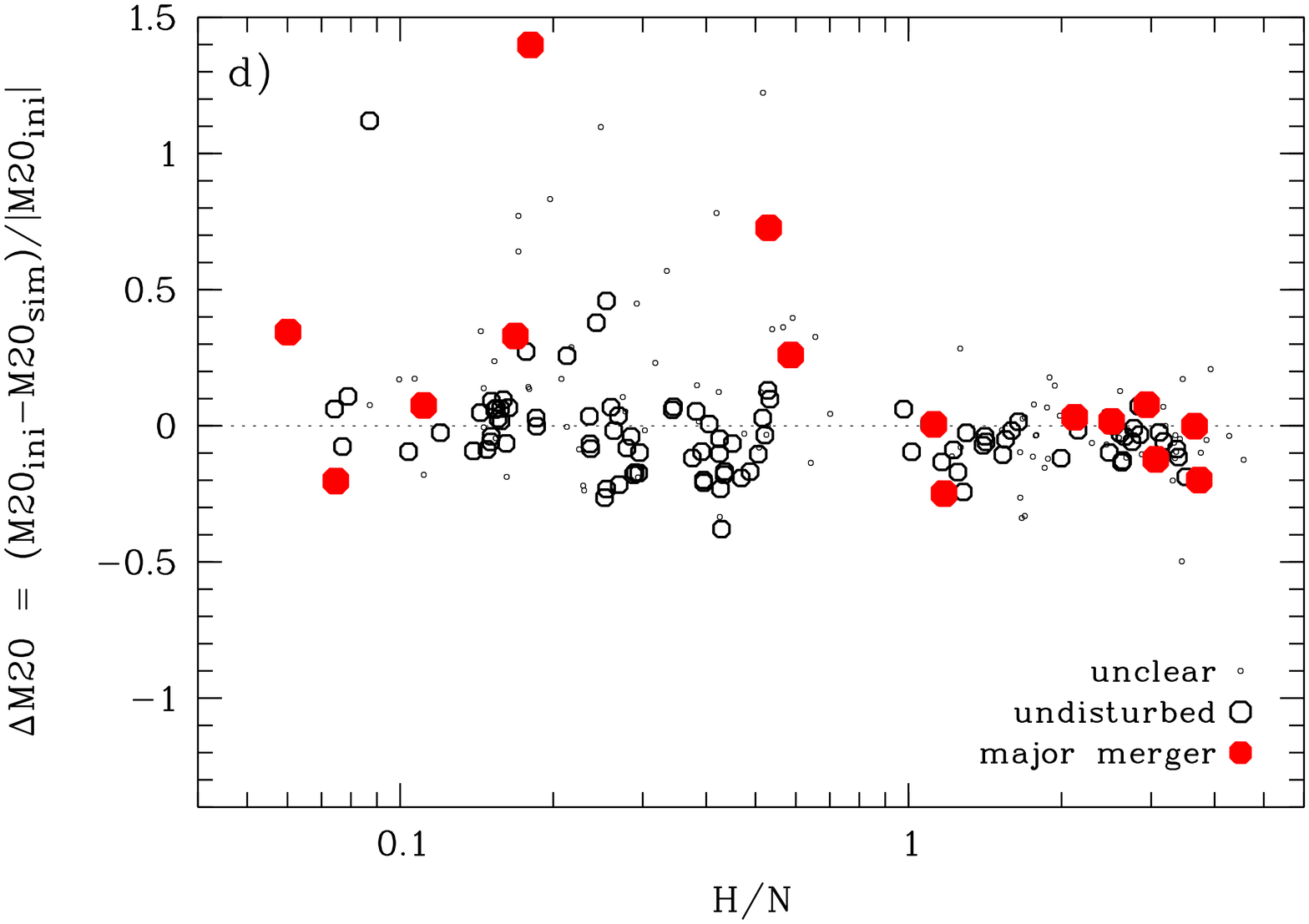}}
\caption{\label{trendsnew2}
Illustration of the impact of a \emph{subtracted} nucleus on the quantitative 
morphologies of the simulated AGN galaxies in concentration index $C$ (upper left),
asymmetry $A$ (upper right), Gini coefficient $G$ (bottom left), and \mtw\ (bottom right).
As a function of $H/N$, all plots show the initial parameter value before (index ini) 
and after addition \& subtraction of a synthetic nucleus (index sim), normalized to the 
initial value. See text for details.
} 
\end{figure*}

%
%
\begin{figure*}
\resizebox{\hsize}{!}{\includegraphics[height=9cm,angle=270]{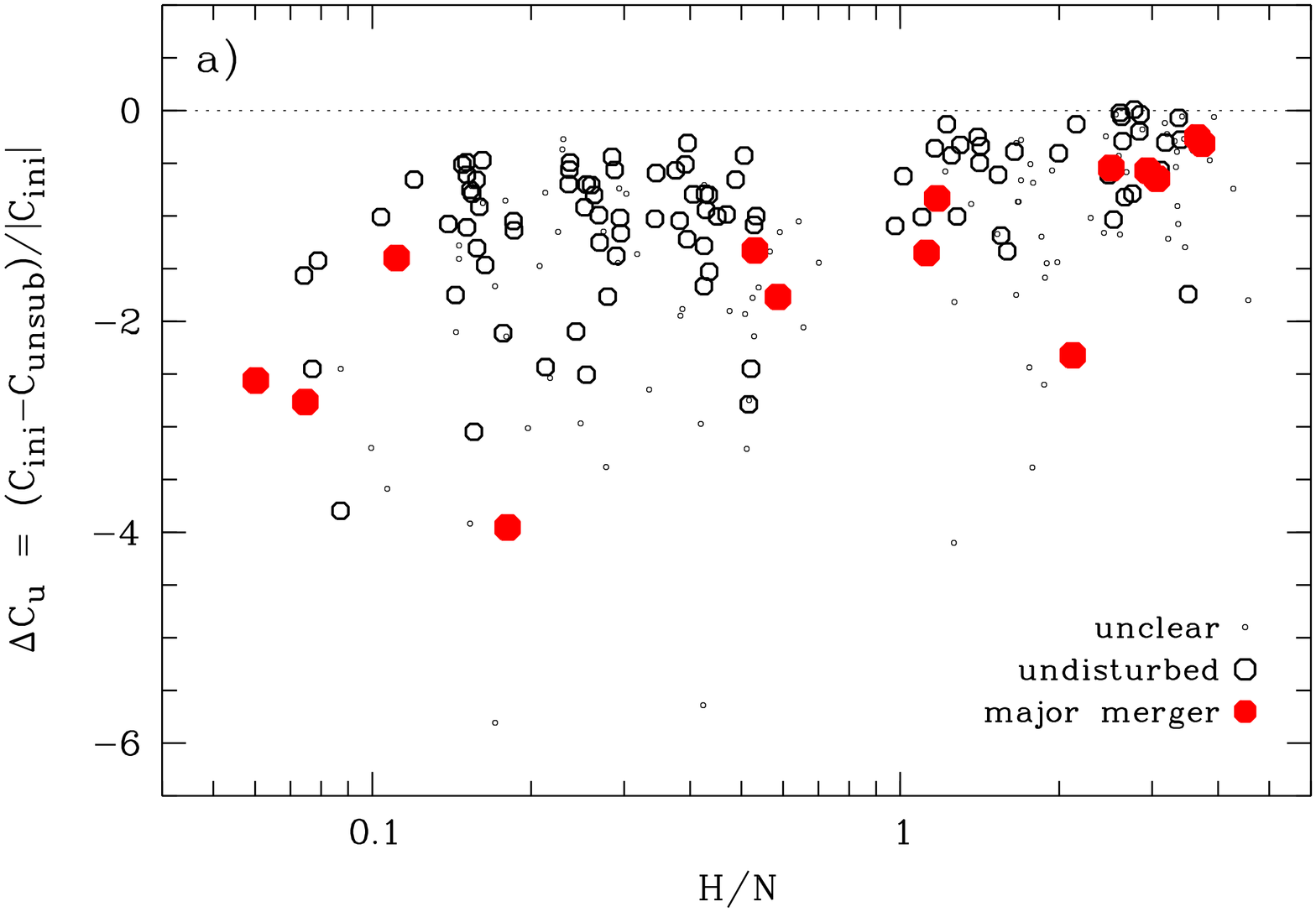}
\includegraphics[height=9cm,angle=270]{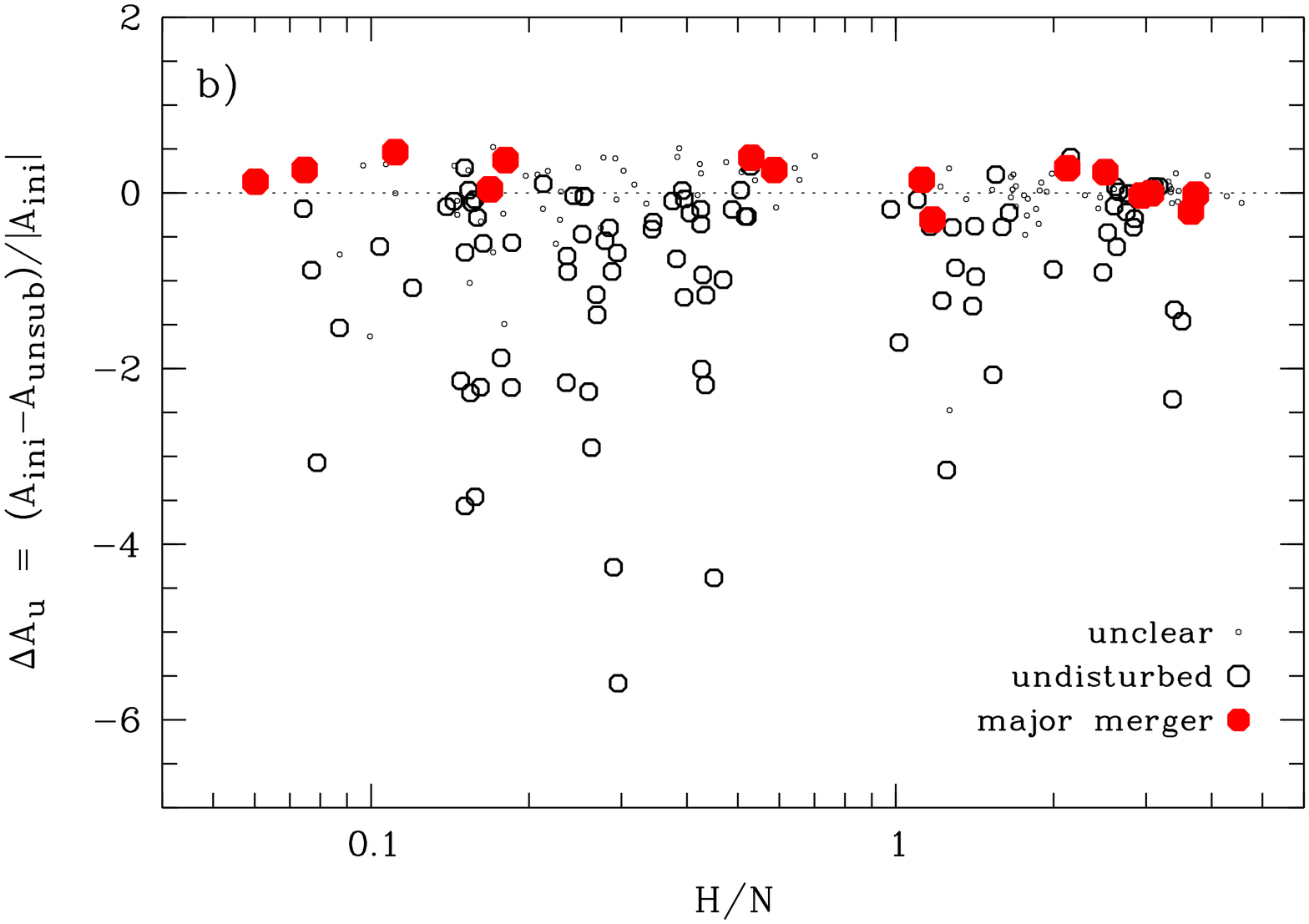}}
\resizebox{\hsize}{!}{\includegraphics[height=9cm,angle=270]{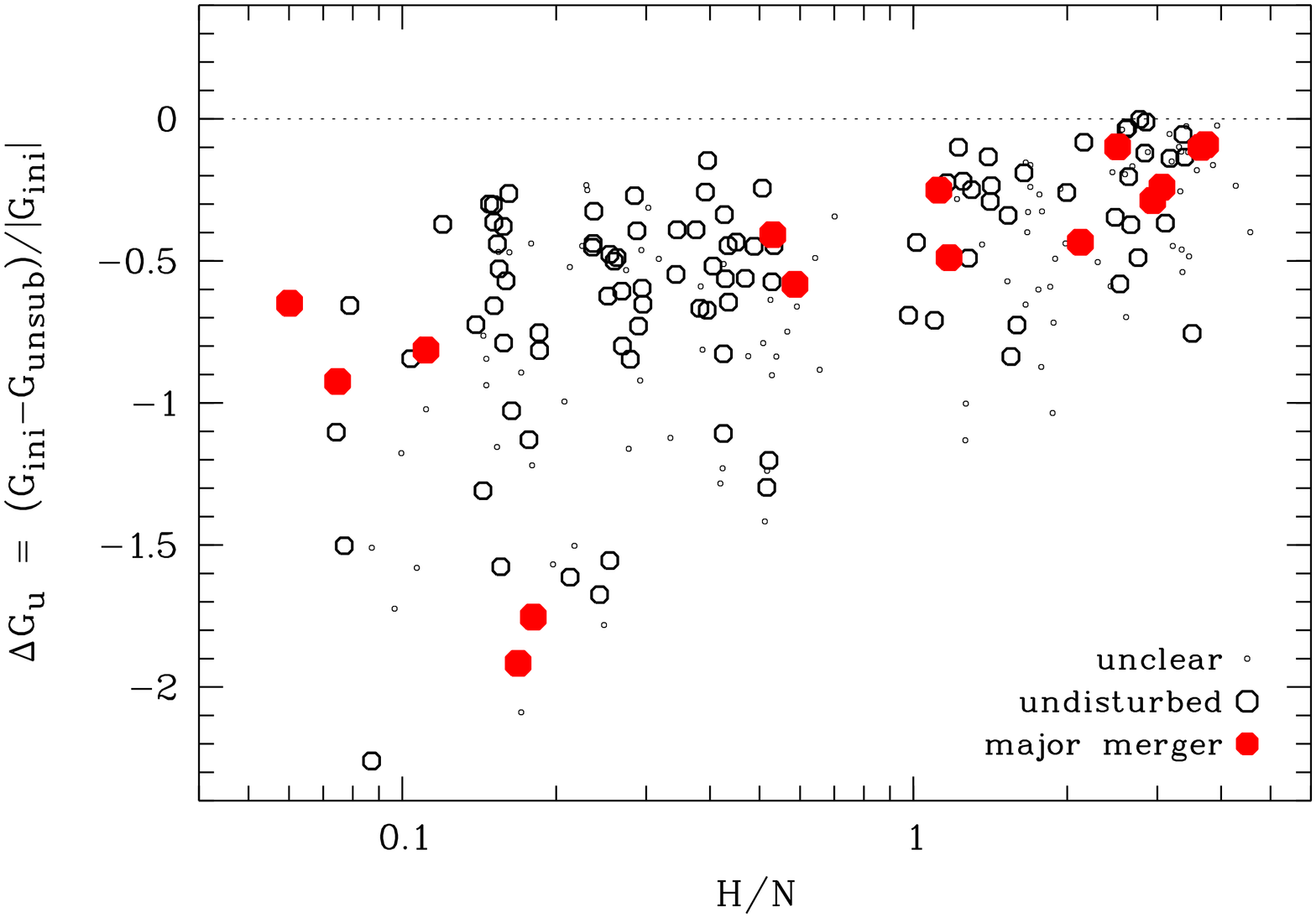}
\includegraphics[height=9cm,angle=270]{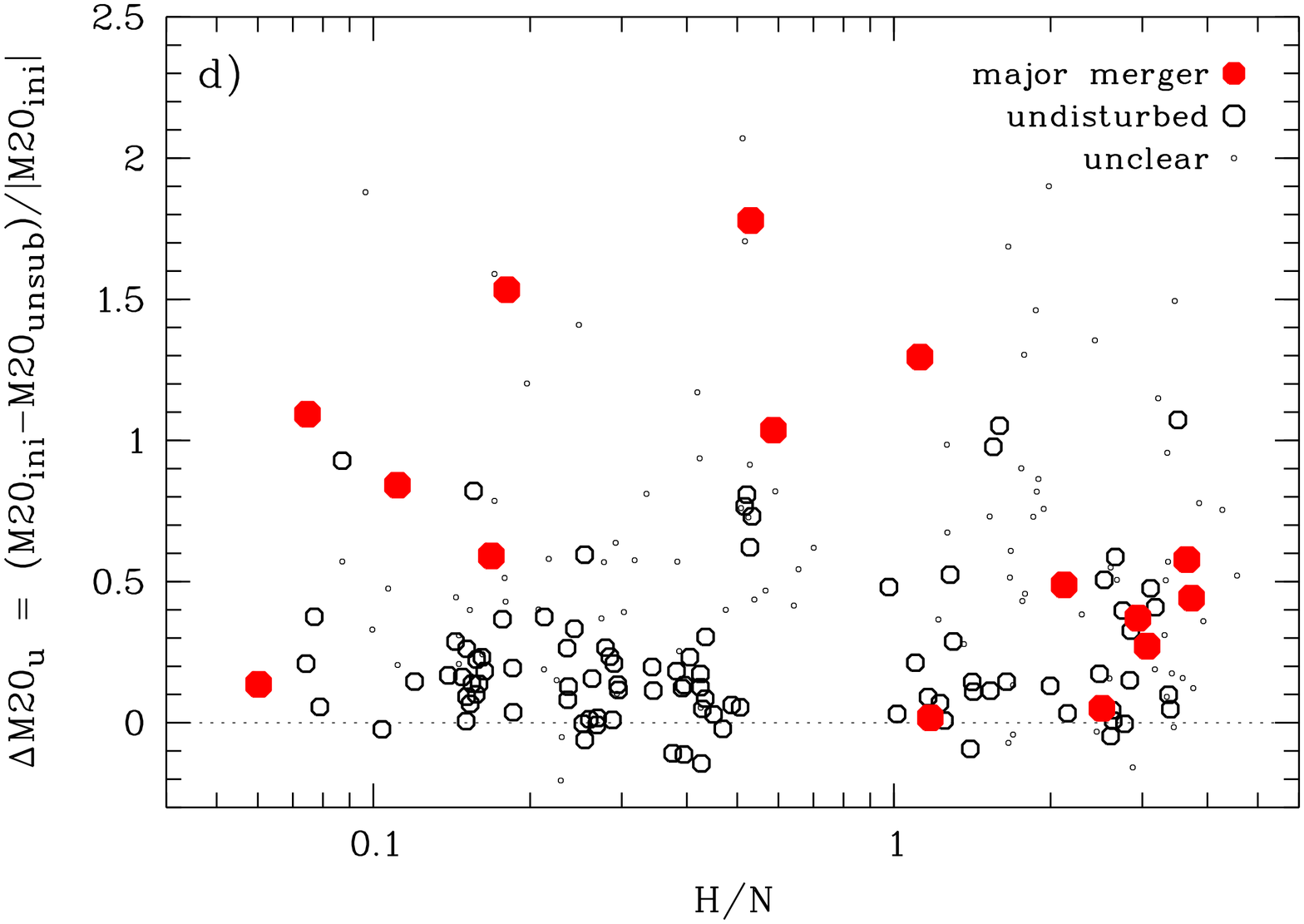}}
\caption{\label{trendsnew3}
Illustration of the impact of an \emph{unsubtracted} nucleus on the quantitative 
morphologies
of the simulated AGN galaxies in concentration index $C$ (upper left),
asymmetry $A$ (upper right), 
Gini coefficient $G$ (bottom left), and \mtw\ (bottom right).
As a function of $H/N$, both plots show the 
initial value before (index ini) and after the addition of a 
synthetic nucleus (index unsub), normalized to the initial value.
See text for a discussion. 
} 
\end{figure*}

The presented distributions in the parameter spaces  spanned by
$C$, $A$, $G$, and \mtw\ already gave some insight into the effects
of an optical nucleus in the morphological analysis.
We now investigate this in more detail for each of the four
descriptors in Fig.~\ref{trendsnew2}.
In the upper left plot,
$C_{\rm ini}$ is the initial concentration index
of a non-AGN galaxy \emph{before} 
adding a synthetic nucleus, $C_{\rm sim}$ gives the index after 
addition and subtraction of a synthetic nucleus. The difference between the two 
is normalized to the absolute initial value via 
$\Delta C = (C_{\rm ini}-C_{\rm sim}/|C_{\rm ini}|)$. This parameterization 
allows us to distinguish between systematic trends, showing up as
changes in the average value of $\Delta C$, and 
a loss of information leading to an increased scatter in $\Delta C$.   
We show $\Delta C$ as a function of $H/N$.

After subtracting the synthetic nucleus, the concentration indices  
$C_{\rm sim}$ are on average slightly underestimated for high $H/N$ 
values, corresponding to $\Delta C>0$. 
This trend is the equivalent of the underestimated 
S\'ersic indices $n_{\rm sim}$ in Fig.~\ref{galfitplots}e. 
When decomposing the AGN into host and nucleus, some part of the
host's flux is systematically attributed to the central point source.
However, this approach introduces fewer systematic effects
than leaving the nucleus unaccounted for, as we will show below.
The scatter in $\Delta C$ becomes larger toward lower $H/N$, 
independent of morphological disturbances. This explains why the 
$C$/$A$ diagram of nucleus-subtracted AGN hosts is more noisy
than that of quiescent galaxies, and the effect is stronger
for objects with lower $H/N$.

The asymmetry index (Fig.~\ref{trendsnew2}b) is the only descriptor that
shows a clear systematic effect:
toward lower $H/N$, $A_{\rm sim}$ is increased, probably because of the artificial 
noise from the residuals after subtracting the nucleus. The scatter in 
$\Delta A$ (hence, the noise in the $C$/$A$ diagram) also increases 
with decreasing $H/N$. This trend is much weaker for major mergers, however~---
mainly due to their high $A_{\rm ini}$ values~---
than for undisturbed galaxies.
 Since we are particularly interested in
correctly identifying major mergers, this supports our results.

The Gini coefficient (Fig.~\ref{trendsnew2}c) shows a trend similar
to $C$, with slightly underestimated values of $G$ at $H/N 
\gtrsim 0.5$ and increasing noise toward lower $H/N$. The scatter 
is slightly smaller for major merger candidates. 
The \mtw\ index (Fig.~\ref{trendsnew2}d) is slightly 
overestimated for high $H/N$ and underestimated for low $H/N$; the noise
increases with decreasing host--to--nucleus contrast, independent of 
the morphology.

Fig.~\ref{trendsnew3} demonstrates how  
the morphological descriptors are affected
when an optical nucleus is \emph{not} accounted for.
Again, we computed the difference 
between the initial value (index ini) and that after 
the addition of a  synthetic nucleus (index unsub) for 
$C$, $A$, $G$, and \mtw. Unsurprisingly, 
the concentration index (upper left graph) 
becomes systematically higher due to the central 
point source, resulting in high negative values of $\Delta C_{\rm u}$.
This effect becomes stronger toward weaker host--to--nucleus contrasts
and affects major mergers stronger than undisturbed galaxies.
The upper right plot shows the change in asymmetry index $A$.
Particularly interesting here is the morphological dependence:
while $A$ on average
becomes larger for undisturbed galaxies after the addition of a nucleus 
($\Delta A_{\rm u}<0$), the asymmetry of major mergers
is slightly \emph{underestimated}. 
Overestimated asymmetry indices probably are caused
by nuclei that are not perfectly symmetrical but 
show a more complex shape of
the PSF. The Gini coefficient shows the same trend as $C$, 
but without a noticeable dependence on morphology.
\mtw\ is systematically underestimated (i.e.~overestimated in absolute
values), 
and this effect is slightly stronger 
for major mergers. The central point sources probably mask out bright
structures that are off-center but still relatively close to the hosts'
centers.

To summarize, we find that an unsubtracted nucleus has 
a strong systematic effect
on all four morphological parameters considered here. In contrast,
subtracting the central point source systematically only affects 
the asymmetry index $A$.
This implies that comparisons between quiescent galaxies and AGN with
non-negligible nuclei will be biased if the impact of the nuclei is ignored.
Similar results have been found e.g.~by Sanchez et al. (2004) or Gabor
et al.~(2009).

\subsection{AGN host colors\label{hostcol}}

We constructed our quiescent galaxy sample in such a way that a fair
comparison to the real AGN hosts is feasible: they are matched in their redshift
and apparent magnitude distributions. A lower limit on the
S\'ersic index was applied to reject bulgeless disks among the non-AGN.
The initial selection criteria did not use any information on spectral
type, except that galaxies classified as AGN within the COMBO-17 survey
did not enter the comparison sample.

For testing purposes, we used the observed color as 
an additional constraint for the definition of the non-AGN sample.
This analysis is restricted to the GEMS part of
our sample, which offers $v$- and $z$-band HST/ACS imaging, while the 
STAGES survey only comprises $v$-band data from HST. 
Note that the 
nucleus-subtracted HST images are our only option to derive 
host colors~--- the ground-based COMBO-17 data have a totally insufficient 
resolution for a host-nucleus decomposition.

Our sample holds 12 real AGN hosts with available GEMS data. Their 
colors span the range $0.86 \le (v-z) \le 2.4$.
Matching the comparison sample to this color distribution reduces it
to a total of 62 galaxies.

We repeated the KS tests using this new non-AGN sample matched in
redshifts, magnitudes, and colors.
Comparing the AGN to major mergers, undisturbed, and unclear objects
in $C/A$ space, we now find
$P_{\rm mm} = 0.04$, 
$P_{\rm undis} = 0.24$ and 
$P_{\rm uncl} = 0.03$, respectively.
As in our initial analysis, 
undisturbed galaxies show the highest probability of
stemming from the same parent distribution as the real AGN hosts. 
In $G$/\mtw\ space, the respective probabilities are
$P_{\rm mm} = 0.29$, 
$P_{\rm undis} = 0.59$ and 
$P_{\rm uncl} = 0.18$. Again, undisturbed galaxies
are most similar to the real AGN hosts. 
Compared to the initial results without a color selection,
the difference in probability between undisturbed and major merger
cases is reduced.

We emphasize that, because of small number statistics, 
these results are not as robust as our initial analysis.
The additional color criterion shrinks the comparison sample of major mergers 
from 15 objects to just six. 
In particular, four very blue major mergers
with colors $(v-z)<0.9$ are excluded from the quiescent sample.
A visual inspection of these galaxies does not reveal
obvious morphological differences to the rest of the major merger cases.
But the very blue ones show relatively low absolute \mtw\ values
and excluding them slightly affects the KS test results in
$G$/\mtw\ space (by increasing the value of $P_{\rm mm}$).

Nevertheless, the reanalysis with a color-constrained comparison sample 
confirms our initial results. The fact that major mergers 
have much bluer average colors ($\langle (v-z) \rangle = 0.95$) than
the real AGN hosts ($\langle (v-z) \rangle = 1.65$)  can
be taken as some spectrophotometric evidence that
major mergers are scarce among the host galaxies.
This agrees with previous findings by, e.g.,
Sanchez et al.~(\cite{san04}) or Schawinski et al.~(\cite{sch07}):
at least up to redshifts $z \approx 1$, AGN hosts show 
intermediate~--- but not very blue~--- colors.

\subsection{Discussion\label{discus}}

Our quantitative morphological analysis shows that the majority of 21 AGN in 
our sample reside in only mildly disturbed or even undisturbed galaxies. 
The same result is found in a visual classification of the
nucleus-subtracted images (see Table~2): Ten of them (48\%) 
have an undisturbed morphology, 11 (52\%) are unclear cases, but
none has been classified as a major merger.
For comparison, the quiescent sample comprises 15 (7\%) major mergers,
95 (47\%) undisturbed and 93 (46\%) unclear objects.

Hence, neither with a quantitative analysis nor
by visual inspection we find evidence that a significant fraction of the
AGN hosts are undergoing major mergers.
This is at variance with a naive expectation for a connection between
major mergers and AGN. 
In principle, there are three 
possible explanations for this: \\
1.~Morphological descriptors like $C/A$ or $G$/\mtw\
might miss 
a significant fraction of strong gravitational interactions on single-orbit ACS
data of intermediate-redshift galaxies; \\
2.~there might be a delay between a major merger and the onset
of the optical AGN phase: the merger signatures could fade and become
undetectable (or very hard to detect) before the nuclear accretion starts,
or \\
3.~violent gravitational interactions 
might not be the main triggering mechanism of nuclear activity in the
redshift regime considered here, which corresponds to look-back
times between $\sim$\,5 and $\sim$\,8\,Gyr. 

Regarding the first of these possibilities, our analysis shows that 
major mergers \emph{should} be detectable at a high success rate,
even in the presence of residuals from a subtracted nucleus.
This is what we conclude from the KS tests between undisturbed and strongly 
interacting galaxies
(probability $P=8.6 \times 10^{-4}$ using $C/A$ or $P=1.1 \times 10^{-3}$ 
using $G$/\mtw, 
cf.~Fig.~\ref{subtracted2} and Table~\ref{kstest2}) 
as well as from the discussion of the trends in
Fig.~\ref{trendsnew2}.
However, even \emph{if} major mergers are triggering the nuclear activity for
a significant fraction of our sample, it is not necessarily the case that 
the morphological imprint of the merger is still detectable 
\emph{by the time the AGN phase has started}. 
The start of the SMBH accretion and the time of the
strongest morphological perturbations (hence the best ``detectability'' of the
merger) might not occur simultaneously.

This brings us to the second option listed above.
Observational evidence for a delay between a
merger and the onset of an AGN phase has been found e.g.~by Schawinski et
al.~(\cite{sch07}) and Bennert et al.~(\cite{ben08}).
Synthetic observations to estimate the time during which a major merger could
be identified have been carried out by
Conselice~(\cite{con06}; using the $CAS$ space) and 
Lotz et al.~(\cite{lot08}; based on $CAS$ and $G$/\mtw). 
The involved time scales
depend  on very many parameters, such as the mass ratios, 
gas and dust content, or the geometry and sense of rotation of the
merging galaxies. According to these authors, 
the phase of the strongest perturbation occurs at the earliest
a few 100\,Myr and at the latest $\gtrsim1$\,Gyr after the
first encounter. However, these analyses did not 
attempt to estimate the time when the nuclear gas accretion starts, 
nor did they
include the impact of an AGN's optical nucleus on the image analysis.

Using numerical simulations, Hopkins et al.~(\cite{hop08}) have
investigated the various evolutionary stages of merger-induced nuclear activity.
The encounter and major merger phases are followed by the coalescence 
and start of the SMBH accretion. 
By the time the object would be observed as an AGN at rest-frame optical wavelengths, 
the tidal features might have faded. For distant objects, the merger signatures 
could be only detectable with very deep observations, 
but not with single-orbit ACS data such as those from GEMS or STAGES.

Concerning the third hypothesis, it can be argued that at least at 
\emph{higher} redshifts and luminosities, there   
is evidence both from observations
and numerical simulations for a connection between
major mergers and AGN phases.
Ultraluminous infrared galaxies (ULIRGs), e.g., which are mostly found
in merging systems (e.g.~Sanders \& Mirabel 1996), have a much higher AGN 
fraction than the overall
population of luminous galaxies (e.g.~Canalizo \& Stockton~\cite{can01}).
Simulations were used to show that major mergers can efficiently transport
gas to the central parts of the merger remnant to feed the central SMBH 
(e.g.~Di Matteo et al.~\cite{dim05}). 
At later stages, the feedback by the AGN can terminate the star formation; this 
scenario might be a necessity to produce the ``red and dead'' ellipticals
observed in the present-day universe (e.g.~Khalatyan et al.~\cite{kha08}).

Studies using very large samples drawn from the Sloan Digital
Sky Survey, on the other hand, 
did not find an increased fraction of interacting
galaxies among AGN hosts at low $z$
(e.g.~Kauffmann et al.~\cite{kau03}, Li et al.~\cite{li08}).
It has to be noted that these local studies on average cover much
lower AGN luminosities than intermediate-redshift samples like
our own. A straightforward comparison is thus not feasible. 
The observational results at differing
 redshifts 
are diverse but indicate that a possible
connection between major mergers and nuclear activity 
seems to weaken toward lower redshifts and lower AGN
luminosities.

Our results are consistent with weak gravitational interactions 
(fly-bys, minor mergers) playing an important role in AGN triggering at 
$z \approx 0.7$. This would also agree with
the increased galaxy density in the vicinity
of quasars (e.g.~Hennawi et al.~\cite{hen06}) and the enhanced AGN fraction
in galaxy pairs at low (e.g.~Ellison et al.~\cite{ell11}) and intermediate
redshifts (e.g.~Silverman et al.~\cite{sil11}).
Tidal interactions or minor mergers 
can induce the transport of gas with low angular momentum to
the central parts of the host to (re-)start SMBH accretion
(e.g.~Hernquist \& Mihos 1995); the host
morphology would be much less affected than in a major merger event.
Indeed we find that  the AGN
hosts are most similar to undisturbed galaxies, 
in the sense of highest KS test probability ($P=0.32$ or $P=0.22$,
depending on which pair of descriptors is used).
Our visually derived fraction of undisturbed AGN hosts is 48\%, and it is
likely that more objects with only weak tidal features
are among the remaining 52\% of unclear cases (Table~\ref{tabagn}).

The lack of signatures for frequent {\it strong} interactions  is 
consistent with the findings at similar redshifts
by Grogin et al.~(\cite{gro05}), 
Pierce et al.~(\cite{pie07}),
Gabor et al.~(\cite{gab09}), 
and Cisternas et al.~(\cite{cis11}).
One major difference between the first two of these
and our analysis is that they
did not account for the potential influence of an optical nucleus. 
The first three of the above samples 
mainly comprise type-2 AGN since they are 
X-ray selected
(Gabor et al.~give a type-1 AGN fraction of $\sim$\,10\%).
Only the data set by Cisternas et al.~(\cite{cis11}) mainly consists
of type-1 AGN (contributing 60\,\%); our own sample is made up entirely of
type-1 AGN.
Despite these differences in the type mix, 
our results and these previous studies agree qualitatively. 
This is to be expected
in the context of the classical unification scheme of AGN 
(e.g.~Urry \& Padovani~\cite{urr95}), where the distinction between
type 1 and type 2 is assumed to be an observational effect
and not an intrinsic property.

To further interpret the observational findings, 
it would be highly desirable to create synthetic
images based on numerical simulations. With simulations that
include galaxy interactions and mergers in various configurations 
\emph{and} a prescription for AGN activity
(as, e.g., Di Matteo et al.~\cite{dim05}), 
sets of synthetic observations would have to be generated
to investigate, for a given data quality,  
the detectability of interaction signatures during the AGN phase. 
Such an approach 
is, however, beyond the scope of this paper.
In the future, we will also aim at developing a
morphological descriptor scheme that is better
capable of tracing weak interaction signatures than
those we currently have at our disposal.

\section{Summary \label{summary}}

We carried out a quantitative comparison between the morphologies of the host
galaxies of AGN and quiescent galaxies at
redshifts $0.5 < z < 1.1$. The imaging data were taken
from the large HST/ACS mosaics of the GEMS and STAGES surveys.
Our main aim was to test whether the AGN host galaxy population shows an 
increased fraction of morphological perturbations with respect to non-AGN. 
We created synthetic images to investigate the impact 
of an optical nucleus on the morphological analysis 
of AGN host galaxies. To quantify galaxy morphologies, we used the asymmetry 
index $A$, the concentration index $C$, the Gini coefficient $G$, and the \mtw\ index. 
A sample of $\sim$\,200 synthetic AGN was matched to 21 real AGN in terms of 
redshift, host brightness and host--to--nucleus ratio to ensure a reliable 
comparison between active and quiescent galaxies. The optical nuclei 
of type-1 AGN strongly affect the quantitative morphological properties of the 
underlying host galaxy. When these effects are accounted for,
active and inactive galaxies show similar distributions in $C/A$ and 
$G/$\mtw\ space. The morphologies of the AGN hosts are 
clearly distinct from galaxies undergoing strong gravitational interactions. 
This is at variance with a naive expectation for a connection between 
major mergers and AGN+.
However, the signatures of gravitational interactions might be too weak to be detected
by the commonly used descriptors like $C/A$ due to a delay between
the strongest morphological perturbation and the onset of the AGN phase.

\begin{acknowledgements}
We thank the anonymous referee for many suggestions that helped to 
improve the manuscript.
AB acknowledges funding by the Deutsches Zentrum f\"ur Luft- und Raumfahrt
(50\,OR\,0404) and by the  Austrian Science Foundation FWF (grants
P19300-N16 and P23946-N16);
MB and EvK acknowledge support by the FWF under 
grant P18416;
MEG and CW are supported by STFC Advanced Fellowships; 
CH by an European Commission Programme $6^{\rm
th}$ framework Marie Cure Outgoing International Fellowship under
contract MOIF-CT-2006-21891, and a CITA National fellowship; 
BH is grateful for support from the Science and Technology 
Facilities Council (STFC).
KJ is supported by the Emmy Noether Programme 
of the Deutsche Forschungsgemeinschaft;
SJ by NASA under LTSA Grant NAG5-13063, NSF grant AST-0607748
and $HST$ grant GO-11082 from STScI, which is operated by AURA, Inc.,
for NASA, under NAS5-26555;
DHM by NASA under LTSA Grant NAG5-13102;
CYP by STScI and NRC-HIA Fellowship programs; 
SFS by the Spanish MEC grants AYA2005-09413-C02-02 and the 
PAI of the Junta de Andaluc\'\i a as research group FQM322.
Support for STAGES was provided by NASA through GO-10395 from STScI operated
by AURA under NAS5-26555.
\end{acknowledgements}

\end{document}